\documentclass[aps,prd,amsmath,floats,floatfix, twocolumn,
superscriptaddress,nofootinbib,showpacs]{revtex4} 
\usepackage{hyperref}
\usepackage{graphicx} 
\usepackage{bm}

\usepackage{ulem} 
\usepackage[usenames]{color}

\usepackage{longtable}

\newcommand{\CITA}{\affiliation{Canadian Institute for Theoretical
    Astrophysics, 60 St.~George Street, University of Toronto,
    Toronto, ON M5S 3H8, Canada}} %

\begin{document}

\title{
Precessing Binary Black Holes Simulations: Quasicircular Initial Data}

\author{Abdul H. Mrou\'e}  \CITA
\author{Harald P. Pfeiffer}  \CITA
 
\date{\today}

\begin{abstract}
 In numerical evolutions of binary black holes (BBH) it is desirable
  to easily control the orbital eccentricity of the BBH, and the
  number of orbits completed by the binary through merger.  This paper
  presents fitting formulae that allow to choose initial-data
  parameters for generic precessing BBH resulting in an orbital
  eccentricity $\sim 10^{-4}$, and that allow to predict the number of
  orbits to merger.   We further demonstrate how these fits can
  be used to choose initial-data parameters of desired non-zero
  eccentricity.  For both usage scenarios, no costly exploratory BBH
  evolutions are necessary, but both usage scenarios retain the
  freedom to refine the fitted parameters further based on the results
  of BBH evolutions.  The presented fitting formulas are based on 729
  BBH configurations which are iteratively reduced to eccentricity
  $\lesssim 10^{-4}$, covering mass-ratios between 1 and 8 and
  spin-magnitude up to $0.5$.  101 of these configurations are evolved
  through the BBH inspiral phase.
\end{abstract}


\maketitle


\section{Introduction}

The next generation of gravitational-wave detectors, such as advanced LIGO, 
VIRGO and KAGRA~\cite{AdvancedLIGOWebsite,LIGOScience,Abadie:2012,Acernese:2008,Somiya:2012}
is under construction. These advanced detectors will have an order of magnitude 
increase in their sensitivity and are widely expected to make the first direct detections of gravitational 
waves by sources such as coalescing compact object binaries.  The ability to detect and study
these systems depends on the quality and accuracy of the theoretical waveform
models used in building search templates for these detectors.

After breakthroughs in numerical relativity~\cite{Pretorius2005c}, direct numerical calculation of the signal 
emitted during the last phase of the binary's life have become available and have
been rapidly improving (e.g.~\cite{Centrella:2010,Pfeiffer:2012pc}).
These numerical waveforms guide analytical modelers and data analysts in 
the construction of analytical templates~\cite{Buonanno2007,
Ajith-Babak-Chen-etal:2007b,Damour2009a,Buonanno:2009qa,Ajith2009,Pan:2009wj,
ninjashort,Santamaria:2010yb,Ajith:2012tt} and are used to assess the properties
of gravitational wave data-analysis pipelines~\cite{Aylott:2009ya,Ajith:2012tt}.

The parameter space for numerical studies of binary black holes is seven dimensional: Mass-ratio, and the two spin-vectors of the two black holes.  The total mass
scales out due to the scale invariance of the vacuum Einstein equations, and
eccentricity is expected to be radiated away during the preceding gravitational-wave driven inspiral~\cite{PetersMathews1963,Peters1964}.  
Moreover, simulations for gravitational wave data-analysis need to cover a large
number of inspiral orbits and have sufficient accuracy~(e.g. the recent review~\cite{Ohme:2011rm}).  These requirements increase the computational cost of BBH simulations and limits the number of simulations that can be performed. 
Therefore, BBH simulations have generally focused on lower-dimensional subspaces of the entire BBH parameter space, for instance non-spinning systems, 
or systems with spins aligned with the angular momentum.
For example, the number of distinct BBH parameters  
configurations used in Ninja-2~\cite{Ajith:2012tt} was 29: 6 configurations with non-spinning black holes, and 20 with aligned spins (some configurations were computed independently by several groups for a total of 40 numerical simulations).
Numerical simulations of non-precessing BBH systems are also much more
carefully studied, e.g.~\cite{Baker2006e,Bruegmann2006,Campanelli2006a,Campanelli2006c,Husa2007,Boyle2007a,Baker-Campanelli-etal:2007,BodeEtAl:2008,Chu2009,Scheel2009,Hannam:2009hh,Lovelace:2010ne,HannamEtAl:2010,Lovelace:2011nu,Buchman:2012dw}.

Furthermore, most of the work done to compute analytical waveform models
focuses on non-precessing binaries~\cite{Hannam:2010,Santamaria:2010yb,Ajith2009,Ajith:2008b,Ajith:2009,Ajith-Babak-Chen-etal:2007b,Tiec:2011bk,Mroue2008,Taracchini:2012,Boyle:2008,PanEtAl:2011}. 
Significantly less work has been done in trying to understand dynamics of precessing 
binaries and the resulting waveforms, and how the numerical data agrees with 
the analytical approximations~\cite{Campanelli2007a,Campanelli2007b,Schmidt2010,OShaughnessy2011,Boyle:2011gg,Sturani:2010ju,Sturani:2010yv,O'Shaughnessy:2012vm}.

The first step for evolving a precessing system requires low-eccentricity
initial data, since the orbit of a isolated binary circularizes during the 
inspiral via the emission of gravitational waves~\cite{Peters1964,PetersMathews1963}. 
Even binaries starting with some eccentricity at the beginning of their evolution 
are expected to have a negligible eccentricity near the merger phase, when the 
emitted signal enters the frequency band of these ground-based detectors.

In a numerical relativity simulation of the inspiral of binary black holes, the eccentricity is 
predetermined by three free parameters in the initial data: the orbital
frequency $\Omega_0$, the separation between the holes $r_0$ and the radial 
velocity $\dot{r}_0$ (we often use the dimensionless expansion factor 
$\dot{a}_0 \equiv \dot{r}_0/r_0$).  When initial data is constructed with the 
assumption of circular orbits, the resulting trajectories have an orbital 
eccentricity of the order one percent~\cite{Buonanno-Cook-Pretorius:2007,
Pfeiffer-Brown-etal:2007,Baker2006d}.
This eccentricity arises by neglecting the small, but non-zero, radial 
velocity and the initial relaxation of the black holes. 

A variety of eccentricity definitions are given in the analytical literature 
~\cite{Lincoln-Will:1990,Damour-Schafer:1988,Damour2004,KonigsdorfferGopakumar2006,Memmesheimer-etal:2004,Berti2006,Will-Mora:2002}
as well as in numerical relativity
~\cite{Buonanno-Cook-Pretorius:2007,Baker2006d,Pfeiffer-Brown-etal:2007,Husa-Hannam-etal:2007,CampanelliEtal2009,Mroue2010,Buonanno:2010yk}.
All of these definitions employ residual oscillations in the orbital variables 
such as the orbital frequency, proper horizon separation and coordinate 
separation to estimate the eccentricity. 
Several methods are introduced to 
 choose initial-data parameters that result in lower eccentricity.
The evolution of post-Newtonian equations was used to find quasi-circular 
parameters for the trajectories of binaries~\cite{Husa-Hannam-etal:2007}. 
More recently, iterative procedures were developed to remove eccentricity 
from the initial 
data~\cite{Pfeiffer-Brown-etal:2007,Boyle2007,Tichy:2010qa,Buonanno:2010yk,Purrer:2012wy}.  
These methods first utilize post-Newtonian (PN) information to find initial data with reasonably 
low eccentricity.
The initial data is evolved for about two to three orbits, and after analyzing
the orbit, the initial-data parameters are corrected to reduce the orbital eccentricity.  
Using these new initial-data parameters, the procedure is repeated until the desired value of eccentricity is 
obtained. For non-precessing binaries, the method using proper horizon 
separation worked well in reducing the eccentricity of 
binaries~\cite{Buchman-etal-in-prep,Chu2009,Lovelace:2010ne}. 
For precessing binaries, a new method was introduced in 
Ref.~\cite{Buonanno:2010yk} which uses the instantaneous orbital frequency
to reduce the eccentricity to below $10^{-4}$ in about four iterations. 

%
Iterative eccentricity removal works
  well~\cite{Buonanno:2010yk}, however, it introduces extra steps into
  the numerical simulation of BBH systems.  The goal of this paper is
  to deal with eccentricity removal for conformally flat BBH systems
  once and for all:  We systematically apply the eccentricity removal
procedure to a large number of different BBH systems of different
mass-ratios, spin-magnitudes and spin-directions.  We cover
mass-ratios between 1 and 8, and spin-magnitudes up to 0.5.  We then
perform fits that will allow us to predict low-eccentricity
initial-data parameter for BBH systems with any spin-orientations.  As
we show, these fits result in initial conditions with remaining
eccentricity of $\sim 10^{-4}$.  This is quite likely sufficient for
all near-term GW data-analysis purposes~\cite{BrownZimmerman2009}.
These low-eccentricity fits can also be used to predict initial data
parameters of a desired non-zero eccentricity.  Finally, we evolve 101
of the 729 low-eccentricity initial data sets through their entire
inspiral, and use this information to prepare a fitting formula that allows
us to predict the number of inspiral orbits for generic BBH binaries.

This paper is organized as follows: In Sec.~\ref{sec:ecc_reduction},
we discuss eccentricity and motivate a new definition of eccentricity,
$e_{\rm R}$, that uses the radial velocity in addition to the orbital
frequency.  Once low-eccentricity parameters are known, $e_R$ allows
us to estimate the eccentricity of any parameter choice nearby without
evolving the initial data.
In Sec.~\ref{sec:ecc_NR}, we quasi-circularize 
iteratively 729 BBH configuration with different mass-ratios, spin-magnitudes and -orientations, and various initial separations.  
Three of these simulations are used to compare our newly defined eccentricity 
$e_{\rm R}$ to the standard eccentricity based on the orbital frequency, 
$e_\Omega$, used in Ref.~\cite{Buonanno:2010yk}. 
Section~\ref{sec:fitting} introduces fitting formulae to the low-eccentricity configurations, and demonstrates their efficiency and their advantages over post-Newtonian formulas.
In Sec.~\ref{sec:eccentricity}, we provide formulas for generating initial-data
parameters that result in evolutions of a predetermined eccentricity, and 
assess the quality of these formulas.
Finally, we summarize our conclusions in Sec.~\ref{sec:conclusions}.

\section{Defining and reducing Eccentricity}
\label{sec:ecc_reduction}

\subsection{Newtonian dynamics}

For two bodies on a  Newtonian orbit with small eccentricity $e$, the 
distance $r(t)$ between their centers and the orbital frequency $\Omega(t)$ 
can be written as
\begin{eqnarray}
\label{eq:e-rN}
r(t) &=& \,\bar r\,\,\left[1-e\, \sin (\bar \Omega t + \phi_0)\right] + {\cal O}(e^2), \\
\label{eq:e-rW}
\Omega(t) &=& \bar \Omega\, \left[1 + 2e\, \sin ( \bar \Omega t + \phi_0)\right] +{\cal O}(e^2)\,.
\end{eqnarray}
Here $\bar r$ is the average separation, $\bar \Omega$ is the average orbital 
frequency and $\phi_0$ is a phase component. Taking a time-derivative of 
Eqs.~\ref{eq:e-rN} and~\ref{eq:e-rW}, we find
\begin{align}
\label{eq:dotrN}
\dot{r}(t) &=-\bar r\,e\,\bar \Omega \, \cos(\bar \Omega t+\phi_0) +{\cal O}(e^2),\\
\label{eq:dotOmegaN}
\dot\Omega(t)&=\;\;2\,e\,\bar\Omega^2 \cos(\bar \Omega t+\phi_0) +{\cal O}(e^2)\,.
\end{align}
Note that the factor $\bar\Omega^2$ is the product of the average orbital frequency $\bar\Omega$ and the frequency of the oscillations in $\Omega(t)$.  In Newtonian gravity, these two frequencies agree.

For {\em circular} orbits, the distance between the Newtonian masses $r_0$ and their orbital frequency $\Omega_0$ are
related by Kepler's law,
\begin{equation}\label{eq:Kepler}
r_0^3\,\Omega_0^2 = Gm
\end{equation}
where $m$ is the total mass of the binary, and $G$ represents Newton's
constant which we will henceforth set equal to unity. In this case, eccentricity $e$ and radial velocity $\dot r(t)$ are both identically zero. 

Let us consider how small perturbations in orbital separation, orbital frequency
and radial velocity will affect the eccentricity of the originally circular
orbit. To accomplish this, we relate a circular orbit with $r_0$ and $\Omega_0$ with a slightly eccentric orbit given by Eqs.~(\ref{eq:e-rN}) and~(\ref{eq:e-rW}).
At time $t=0$, we set $r(0)=r_0+\Delta t$, $\Omega(0)=\Omega_0+\Delta \Omega$,
and we set a radial velocity $\dot r(0)=\Delta \dot r$.
The perturbed orbit will in general have an average distance $\bar r$ different from $r_0$, and an average 
orbital frequency $\bar\Omega$ different from $\Omega_0$.  We write 
\begin{equation}\label{eq:OrbitalPerturbation}
\bar r=r_0+\delta\bar r,\quad \bar\Omega=\Omega_0+\delta\bar\Omega.
\end{equation}
Kepler's third law Eq.~(\ref{eq:Kepler}) is also valid for $\bar r$ and $\bar\Omega$, and from it follows 
\begin{equation}
\frac{\delta \bar r}{ r_0}=-\frac{2}{3}\frac{\delta \bar \Omega}{\; \Omega_0} \equiv \epsilon \,.
\end{equation}
The change $\delta\bar r$ and $\delta\bar\Omega$ are thus determined by one small parameter $\epsilon$, which is as of yet undetermined. To proceed we substitute Eq.~(\ref{eq:OrbitalPerturbation}) into Eqs.~(\ref{eq:e-rN})--(\ref{eq:dotrN}), evaluate these expressions at $t=0$, and equate to the assumed perturbations:
\begin{eqnarray}
r_0 +\Delta r &=&  (1+\epsilon) r_0 (1-e \sin \phi_0) r_0 \,, \\
\label{eq:e-int-2}
\Omega_0 + \Delta \Omega &=&  (1-3 \epsilon/2) (1+2 e \sin \phi_0)
\Omega_0 \,, \\ 
\label{eq:e-int-3}
\Delta\dot r  &=& - (1+\epsilon) r_0 (1-3/2 \epsilon) \Omega_0 e \cos \phi_0\,.  
\end{eqnarray}
To first order in $\epsilon$ and $e$, these equations simplify to
\begin{eqnarray}
\label{eq:temp1}
\frac{\Delta r}{r_0} &=& \epsilon - e \sin \phi_0 \,, \\
\frac{\Delta \Omega}{\Omega_0} &=& -\frac{3}{2}\epsilon + 2 e \sin \phi_0 \,, \\
\label{eq:temp3}
\Delta\dot r &=& -r_0 \, \Omega_0 \,\,e\cos \phi_0.
\end{eqnarray}
Inverting Eqs.~(\ref{eq:temp1})--(\ref{eq:temp3}) yields
\begin{align}
\label{eq:new-eNewt}
e & = \sqrt{\left(2 \frac{\Delta \Omega}{\Omega_0} +3\frac{\Delta r}{r_0}\right)^2 + \left(\frac{\Delta\dot r}{r_0 \Omega_0}\right)^2}\,,\\
\label{eq:new-epsNewt}
\epsilon&=2\frac{\Delta\Omega}{\Omega_0}+4\frac{\Delta r}{r_0}\,,\\
\label{eq:new-phi0Newt}
\tan\phi_0 &= -\frac{3\Omega_0\Delta r + 2 r_0\Delta\Omega }{\Delta\dot r}\,.
\end{align} 
To summarize, perturbing a circular Newtonian orbit by $\Delta r$,
$\Delta\Omega$ and $\Delta\dot r$ results in an orbit with
eccentricity (\ref{eq:new-eNewt}); the perturbed orbit has an average
radius larger by $\delta \bar r/r=\epsilon$ relative to the the circular,
unperturbed orbit.

\subsection{Numerical relativity}

Let us now consider general relativistic BBH binaries computed by numerical 
simulations.  As in earlier
work\cite{Buonanno:2010yk,Mroue2010} we define eccentricity based on
periodic oscillations of the separation of the binary or the orbital
frequency, carrying over the Newtonian definitions.   Specifically, the 
instantaneous orbital 
frequency $\Omega(t)$ and the distance $r(t)$ are calculated from the 
coordinate motion of the apparent horizons' centers. We define their relative 
separation vector $\mathbf{r}(t)= \mathbf{c}_1(t)-\mathbf{c}_2(t)$ with magnitude 
$r(t)=|\mathbf{r}(t)|$, where $\mathbf {c}_{i}(t)$ are the coordinates of the 
center of each black hole. Using standard Euclidean vector calculus, the 
instantaneous orbital frequency is then computed as 
\begin{equation}
\mathbf{\Omega}(t) = \frac{\mathbf{r}(t) \times{\mathbf{\dot r}(t)}}{r^2(t)},
\end{equation}
where $\Omega(t)$ is its magnitude. 

%
%
A compact binary inspiral starts at $t=0$ with an initial separation $r_0$, an 
orbital frequency $\Omega_0$ and a radial velocity $\dot r_0$. The time derivative 
of the orbital frequency $\dot\Omega(t)$ is computed and fitted  with the functional
form
\begin{equation}
\label{eq:dotOmegaFit}
\dot{\Omega}_{\rm NR}(t) =S_{\Omega,{\rm fit}}(t) + B_\Omega\, \cos( \Omega_r t+ \phi ).
\end{equation}
The first part of the fit, $S_{\Omega,{\rm fit}}(t)$, is a non-oscillatory function
that captures the radiation-reaction driven inspiral. The second part captures 
the oscillatory contribution of the orbital eccentricity using fitting 
parameters $B_\Omega$, $\Omega_r$ and $\phi$.  
Note that the sinusoidal frequency in Eq.~\ref{eq:dotOmegaFit} is given by the radial
frequency $\Omega_r$. In general relativity, the orbital frequency $\Omega$ exceeds the
radial frequency $\Omega_r$ causing periastron advance~\cite{Mroue2010}. 
With the fitted parameter $B_\Omega$ equal to $2 e \Omega \Omega_r$ (cf. the comment after Eq.~\ref{eq:dotOmegaN}), we can derive 
updating formulas based on $\dot\Omega(t)$~\cite{Buonanno:2010yk}:
\begin{eqnarray}
\label{eq:adotUpdate-Omega}
\dot a_0 &\to& \dot a_0+ \frac{B_\Omega}{2\Omega_0}\,\cos\phi,\\
\label{eq:OmegaUpdate-Omega}
\Omega_0&\to&\Omega_0-\frac{B_\Omega \Omega_r}{4\Omega_0^2}\,\sin\phi\,,
\end{eqnarray}
where $B_\Omega$ and $\phi$ are fitted for while $\Omega_0$ and $r_0$ are given
by the initial data. In Eq.~(\ref{eq:adotUpdate-Omega}), we have introduced the expansion factor $\dot{a}_0 = \dot{r}_0/r_0$ which  appears
naturally in our formulation of the initial-value problem for
BBH binaries with radial velocity~\cite{Pfeiffer-Brown-etal:2007}.
 We use Eqs.~(\ref{eq:adotUpdate-Omega}) and~(\ref{eq:OmegaUpdate-Omega}) 
 to reduce eccentricity 
iteratively for every simulation in this paper.
Using the fitted $B_\Omega$ and $\Omega_r$ in addition to $\Omega_0$,
we estimate the eccentricity $e_\Omega$ as 
\begin{equation}
e_\Omega=\frac{B_\Omega}{2 \Omega_0 \Omega_r}
\end{equation}

Equation~(\ref{eq:new-eNewt}) gives the eccentricity of a Newtonian
orbit that differs by $\Delta\Omega$ and $\Delta\dot r$ from being
circular.  If we know non-eccentric initial data parameters
$\Omega_{0,e=0}$ and $\dot a_{0,e=0}$, we can use the deviations from
the $e=0$ parameters to define eccentricity:
\begin{equation}
\label{eq:e-R}
e_{\rm R} \equiv \left[ \left(\frac{\dot a_0 -\dot a_{0,e=0}}{ \Omega_0} \right)^2 + \left(2\frac{\Omega_0 - \Omega_{0,e=0} }{\Omega_c} \right)^2 \right]^{1/2}\!\!.
\end{equation}
We will test this formula in Sec.~\ref{sec:ecc_NR}, and use it in
Sec.~\ref{sec:eccentricity} to propose a technique to construct BBH
initial-data with specified non-zero eccentricity.

\section{Low-eccentricity BBH parameters}
\label{sec:ecc_NR}

Our goal is to compute several hundred low eccentricity
initial-data sets for subsequent evolutions.  We proceed in two
stages: In stage 1, we circularize 69 binaries by starting with
post-Newtonian estimates for the initial orbital frequency and
radial velocity.  We perform three to four levels of iterative
eccentricity removal.  We use the low-eccentricity initial data
parameters to generate fitting formulae that predict
low-eccentricity initial data parameters for other choices of masses
and spins.  In stage 2, we use these fitting formulae to compute 660
additional BBH configurations, which we find have an average
eccentricity of about 0.003, i.e. a factor of $\sim 5$ below the
first stage simulations.  We finally apply iterative eccentricity
removal to the stage 2 simulations until the eccentricity is less than
0.001.

\subsection{Numerical methods}

We construct BH--BH initial data using the conformal thin sandwich
formalism~\cite{York1999,Pfeiffer2003b} with quasi-equilibrium
boundary conditions~\cite{Cook2002,Cook2004,Caudill-etal:2006} and a
radial velocity as in Ref.~\cite{Pfeiffer-Brown-etal:2007}.  All
initial-data sets considered here use conformal flatness and maximal
slicing. The resulting set of five nonlinear coupled elliptic
equations is solved with multi-domain pseudo-spectral techniques
described in Ref.~\cite{Pfeiffer2003}
implemented in the Spectral Einstein Code {\tt SpEC} \cite{SpECwebsite}.
Calculation of initial data
with desired physical parameters (masses $m_A, m_B$ and dimensionless
spins $\vec \chi_A$, $\vec\chi_B$) requires a root-finding procedure
to determine the corresponding initial data parameters (radii of the
excision boundaries $r_A$, $r_B$ and angular velocities of the
horizons, $\vec\Omega_A$ and $\vec\Omega_B$).  This root-finding is
described in Ref.~\cite{Buchman:2012dw}.  

The initial data is evolved with {\tt SpEC} \cite{SpECwebsite}, using
a first-order representation~\cite{Lindblom2006} of the generalized
harmonic system~\cite{Friedrich1985,Garfinkle2002,Pretorius2005c} that
includes constraints damping
terms~\cite{Gundlach2005,Pretorius2005c,Lindblom2006}.  Constraint
damping parameters are chosen as in~\cite{Buchman:2012dw}, based on
experience gathered in~\cite{Chu2009}.  The computational domain
extends from excision boundaries located just inside each apparent
horizon to some large radius, where the outgoing gravitational
radiation pass freely through the outer boundary.
Outer boundary conditions~\cite{Lindblom2006,Rinne2006,Rinne2007} are 
imposed to prevent the influx of constraint
violations~\cite{Stewart1998,FriedrichNagy1999,Bardeen2002,Szilagyi2002,%
Calabrese2003,Szilagyi2003,Kidder2005} and undesired incoming
gravitational radiation~\cite{Buchman2006,Buchman2007}, 
while no boundary
conditions are imposed at the inner excision boundaries.
Interdomain boundary conditions are enforced with a penalty
method~\cite{Gottlieb2001,Hesthaven2000}.  The overall evolution
techniques (constraint damping parameters, choice of domain
decomposition) are essentially identical to the inspiral phase of
Ref.~\cite{Buchman:2012dw}.  For precessing runs, we employ a
coordinate transformation based on pitch- and yaw-angles (i.e. two of
the Euler angles), described in detail in a forthcoming
publication~\cite{Ossokine-etal-in-prep}.  This technique works well
for moderate precession as in the cases presented here, although 
in future runs, it will be replaced by the more sophisticated
coordinate transformations developed in~\cite{Ossokine-etal-in-prep}.

\begin{figure}
\includegraphics[scale=0.47]{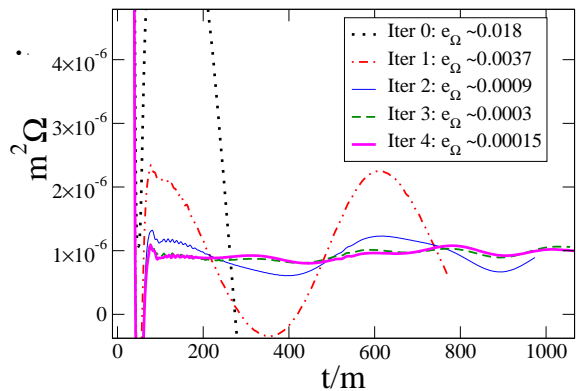}
\caption{
\label{fig:ecc-removal-q1.5} 
{\bf Eccentricity removal based on time derivative of the orbital 
frequency $d\Omega/dt$}, applied 
to a precessing binary black hole, ScN 21. Shown is $\dot\Omega$
vs. time and four eccentricity-removal iterations.
}
\end{figure}

\subsection{Iterative Eccentricity Reduction}
\label{sec:master_run}

We shall start by describing eccentricity removal for one
typical precessing configuration, labeled ScN 21.  This
configuration has a mass-ratio of $q=m_A/m_B=1.5$ and both black holes
have dimensionless spins $0.5$ initially tangent to the orbital plane.
The spin of the larger black hole points exactly away from the smaller 
black hole, whereas the spin of the smaller black hole is anti-aligned 
with its initial velocity. The initial coordinate separation between 
the holes is $r\!=\!16m$.
We begin eccentricity removal with orbital parameters determined from
a non-spinning PN approximant to choose
$\Omega_0\!=\!0.014427/m$, and we use a previous good approximation
for $\dot a_0\!=\!-3.6\times 10^{-5}$ from other simulations.

We evolve the binary with the initial orbital parameters
$(\Omega_0, \dot a_0)$ for about two orbits, and record the
time-derivative of the orbital frequency, $\dot\Omega(t)$.  This
data is plotted in Fig.~\ref{fig:ecc-removal-q1.5} as ``Iter 0''
and we deduce an eccentricity of $e_\Omega\sim 0.018$.
Equations~(\ref{eq:adotUpdate-Omega})
and~(\ref{eq:OmegaUpdate-Omega}) give the improved values for
$\Omega_0$ and $\dot a_0$ to use in next iteration, labeled ``Iter
1'' in Fig.~\ref{fig:ecc-removal-q1.5} with eccentricity
$e_\Omega=0.0037$.  This procedure is repeated three more times
until a final eccentricity $e_\Omega\sim1.5\times10^{-4}$ is
achieved.  Note that as the eccentricity falls below $e\lesssim
0.001$, spin-induced oscillations become apparent at twice the
orbital frequency, as discussed in
Ref.\cite{Buonanno:2010yk}.  The spin-induced  oscillation dominate
for $e\lesssim 0.0003$, making unambiguous determination of a residual
orbital eccentricity difficult.

\begin{figure}
\includegraphics[scale=0.47]{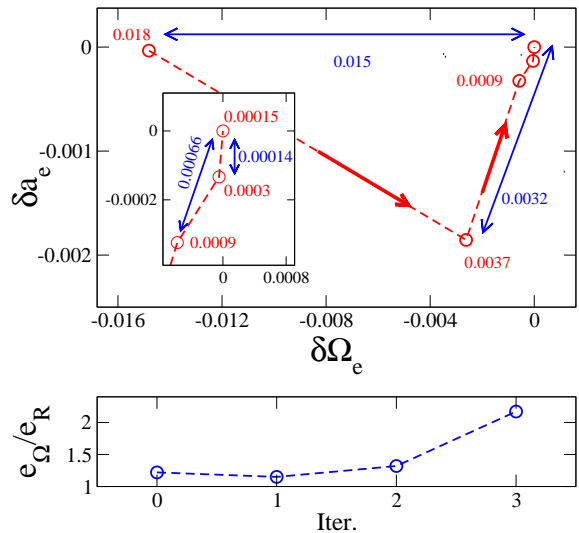}
\caption{
\label{fig:ecc-removal-eR-convergence} 
{\bf Convergence of the eccentricity-removal procedures in the 
($\delta \Omega_e$, $\delta a_e$)  plane.} 
In the upper panel, we show the eccentricity removal sequence of 
Fig.~\ref{fig:ecc-removal-q1.5}.
The red number next to each point gives the eccentricity $e_\Omega$, while
the blue number next to each blue solid line gives the relative eccentricity 
$e_{\rm R}$.
In the lower panel, we plot the eccentricities ratio $e_\Omega/e_{\rm R}$. 
}
\end{figure}

Equation~(\ref{eq:e-R}) indicates that the eccentricity is the square
sum of two components, with one measuring the needed change in orbital
frequency and the other the change in radial velocity (or derivative of expansion
factor).  When plotting parameters during eccentricity removal,
Eq.~(\ref{eq:e-R}) suggests a natural scaling of the axes, relative to
the configuration with lowest eccentricity:
\begin{align}
\label{eq:delta-Omega-e}
\delta\Omega_e & \equiv 2\frac{\Omega_0-\Omega_{0,e=0}}{\Omega_0},\\
\label{eq:delta-adot-e}
\delta\dot a_{e}&\equiv \frac{\dot a_0 - \dot a_{0,e=0}}{\Omega_0}.
\end{align}
In these axes, the Euclidean distance to the origin should correspond directly
to the eccentricity, cf. Eq.~(\ref{eq:new-eNewt}).
Figure~\ref{fig:ecc-removal-eR-convergence} shows the sequence of the 
eccentricity reduction iterations shown in Fig.~\ref{fig:ecc-removal-q1.5} plotted 
in these coordinates. The upper panel of the plot 
shows $\delta\Omega_e$ and $\delta\dot a_e$ as the binary is quasicircularized 
iteratively. The distance in the ($\delta\dot a_e,\delta\Omega_e$) plane between any 
point and the origin corresponding to ``Iter 4'' is equal to the eccentricity 
$e_{\rm R}$. Because Iter 4 is not exactly at zero eccentricity, 
this distance is only an approximation to Eq.~(\ref{eq:new-eNewt}), thus
explaining the raise of $e_\Omega/e_R$ at iterations 2 and 3 in the lower panel of Fig.~\ref{fig:ecc-removal-eR-convergence}.

\subsection{Binaries in data set ${\cal S}_0$}

\begin{figure}
\centerline{\includegraphics[scale=0.9]{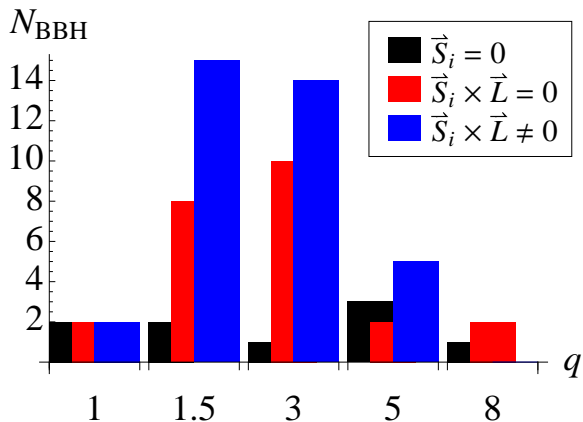}}
\caption{
\label{fig:BarChart} 
{\bf Distribution of configurations in set ${\cal S}_0$.}
Given are the number of non-spinning (black), spinning 
but non-precessing (red) and precessing binaries (blue) for 
different mass-ratio $q$. 
}
\end{figure}

 As a first step, we quasi-circularize iteratively a set of 69
  different configurations, which we shall refer to as set ${\cal
    S}_0$.  The binaries in ${\cal S}_0$ have mass-ratios $1\le q\le
  8$.  There are 9 non-spinning binaries, 24 spinning binaries with
  aligned spins (i.e. without precession), and 36 precessing binaries.
  The dimensionless black hole spin is generally 0.5, but sometimes
  smaller.  The parameters for all 69 configurations are given the
  first nine columns of Table~\ref{tab:Runs} in the Appendix, and
  Fig.~\ref{fig:BarChart} show the distribution of
  non-spinning/aligned spin/non-aligned spins as a function of mass
  ratio: Most precessing binaries have mass ratio either 1.5 or 3.  At
  mass ratio 8, no precessing binaries are evolved.

\begin{figure}
\includegraphics[scale=0.47]{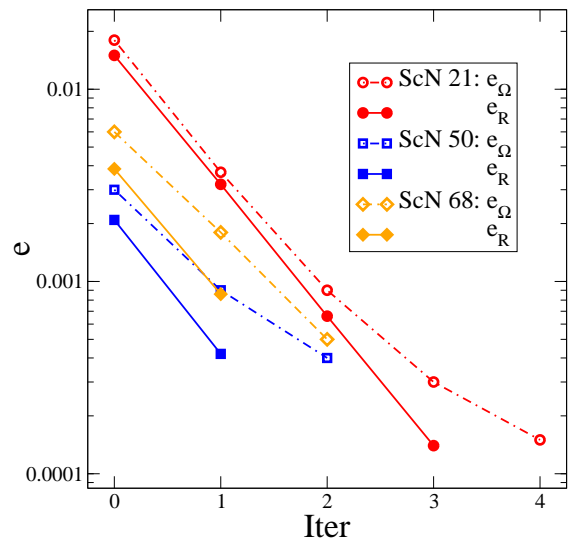}
\caption{\label{fig:ecc-iter-conv} {\bf Convergence of eccentricity 
estimates $e_\Omega$ and $e_R$ for cases ScN 21, 50 and 68.} 
Both eccentricity definitions decrease at the same rate to a value
below the $10^{-3}$ limit. The parameters of the three 
configurations are listed in Table~\ref{tab:Runs}.}
\end{figure}

We apply the iterative eccentricity reduction method to all binaries
of set ${\cal S}_0$.  
Figure~\ref{fig:ecc-iter-conv} illustrates performance of the iterative
eccentricity removal for three cases.  This figure compares also the 
eccentricities $e_\Omega$ and $e_{\rm R}$.
Both eccentricities decrease at the same rate, but the estimated $e_R$
is always smaller than $e_\Omega$, since it is estimated relative to an 
iteration with finite, albeit small, eccentricity. 
As we discuss below, $e_R$ is a useful measure to estimate the eccentricity of runs with the same physical parameters but differing orbital parameters $\Omega_0$ and $\dot{a}_0$,  once the corresponding quasicircularized orbital parameters $\Omega_{0,e=0}$ and $\dot a_{0,e=0}$ are known.

Application of iterative eccentricity removal to all binaries in
${\cal S}_0$ results in orbital parameters $\Omega_0$ and $\dot a_0$
as listed in Table~\ref{tab:Runs}, with an estimated orbital
eccentricity $e_\Omega$.  Eccentricity removal is terminated once an
eccentricity $e_\Omega<10^{-3}$ is reached except for few runs that
are intentionally left with a larger eccentricity for evolutions to
compute periastron advance (cf.~\cite{Tiec:2011bk}).
The initial and final eccentricities of binaries in set ${\cal S}_0$ are
shown in Fig.~\ref{fig:eccentricity-S0}.  The red 
points show the initial eccentricity while the blue point correspond to
the final eccentricity. 

The initial eccentricities for most of these runs
are larger than 0.01. The initial conditions for 
these runs used PN approximants TaylorT3 to predict the values of the 
orbital frequency at a given separation, and the radial scaled velocity is 
set to a small constant value ($\sim-3\times10^{-5}$).
However, for a few runs, the initial eccentricities are much smaller, 
almost $10^{-3}$. 
For these simulations, we used Kepler's law to predict the initial
orbital frequency given the quasi-circular parameters of a closely
similar configuration.  The procedure based on Kepler's law was
experimental and motivated the development of more rigorous fitting
formulas to predict the initial parameters of quasi-circular
binaries, as presented in later parts of this paper. 

All quasi-circular binaries in ${\cal S}_0$ were evolved
  through the inspiral phase until approximately $1-2$ orbits before
  merger.  To provide the reader some context about the length of
  these simulations, Table~\ref{tab:Runs} lists also the number of
  evolved orbits up to a orbital frequency $M\Omega=0.05$.  For
  informational purposes, we also report in Table~\ref{tab:Runs} the
  orbital frequency $m\Omega_f$ at which the simple inspiral evolution
  techniques failed and the number of orbits $N_f$ completed by then.
  Seven binaries were evolved for more than $30$ orbits.  The
  evolutions of merger and ringdown of these binaries require
  significant extension and refinements of previous spectral
  merger-techniques~\cite{Szilagyi:2009qz,Buchman:2012dw}, and will be
  reported elsewhere.

\subsection{Binary configurations $\mathbf{\cal S}_1,\,\ldots,\,{\cal S}_5$}

\begin{figure}
\includegraphics[scale=0.51]{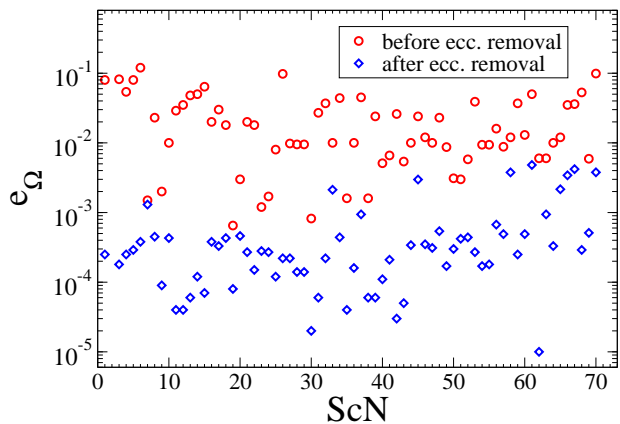}
\caption{\label{fig:eccentricity-S0} {\bf Initial and final
    eccentricities of the configurations in set \boldmath ${\cal
      S}_0$.}  The eccentricity reduction target was $10^{-3}$, except
  for runs intentionally left at larger eccentricity for calculations
  of periastron-advance.  }
\end{figure}

As Fig.~\ref{fig:eccentricity-S0} demonstrates, iterative eccentricity
removal by brute force is possible.  However, even for some
configurations of Fig.~\ref{fig:eccentricity-S0}, the use of ad hoc
fitting formulas based on Kepler's law resulted in significantly
improved initial guesses $\Omega_0$ and $\dot a_0$.  This allowed
to begin iterative
eccentricity removal at smaller initial eccentricity, reducing the
number of eccentricity-removal iterations and thus lowering 
the computational cost of iterative eccentricity removal.
The goal of this section, therefore, is to construct a much larger
sample of low-eccentricity orbital configurations, which will be used
in Sec. IV to develop fitting formulae valid for generic BBH systems.

\begin{figure}
\includegraphics[scale=0.5]{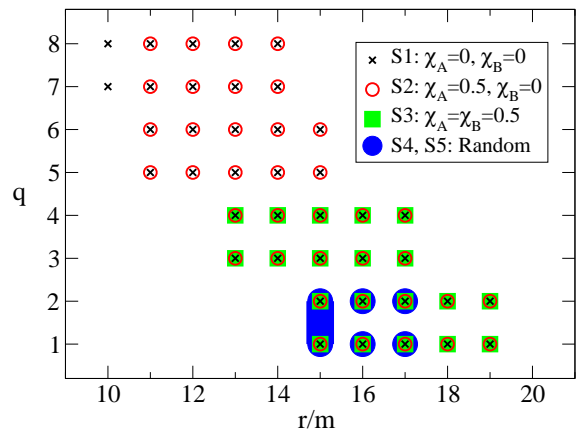}
\caption{\label{fig:ParameterSpace}
Parameter space covered with eccentricity removal.
Each point in set ${\cal S}_1$ represents one configuration for a total of 40. 
Each point in set ${\cal S}_2$ represents five different $\vec\chi_A$ directions, for a total of 190.
Each point in set ${\cal S}_3$ represents 15 different $\vec\chi_A$ and $\vec\chi_B$ directions, for a total of 300.  The set ${\cal S}_4$ and ${\cal S}_5$ contain 130 configurations with randomly generated spin directions and (for 32 of these) randomly generated $q$.}  
\end{figure}

In addition to the 69 configurations of ${\cal S}_0$, we circularized
a total of further 660 configurations, of which 156 are
non-precessing (of which 40 are non-spinning) and 504 are precessing.
The configurations fall into five sets, as indicated in
Fig.~\ref{fig:ParameterSpace}:
\begin{itemize}
\item {\bf Set $\mathbf{\cal S}_1$} contains 40 non-spinning BBH
  configurations. 
\item {\bf Set $\mathbf{\cal S}_2$} contains 190 single-spin
  configurations where the more massive black hole carries spin
  $\chi_A=0.5$.  For each $(q, r/m)$ pair, five different
  spin-directions are considered, with $\theta_A=0, \pi/4, \pi/2,
  3\pi/4,\pi$.  The first and the last of these $\theta_A$ represent
  spins aligned and anti-aligned with the orbital angular momentum, so
  that $76$ configurations are non-precessing, while the remaining
  $114$ configurations are precessing.
\item {\bf Set $\mathbf{\cal S}_3$} contains 300 configurations where
  both black holes carry spin $\chi_A=\chi_B=0.5$.  We consider the same 
five spin directions for the more massive black hole as in set ${\cal S}_2$. 
For each of these, we consider three spin directions for the less massive black hole: $\theta_B=0, \pi/4, \pi/2$.   40 of the configurations in set ${\cal S}_3$ are non-precessing, and the remaining 260 are precessing. 
\item {\bf Set $\mathbf{\cal S}_4$} contains 98 binaries with
  spin-magnitudes $\chi_A=\chi_B=0.5$, and random spin-directions.
  25 binaries were selected for each combination of mass-ratio $q=1$ or $q=2$, 
  and of initial separation $r/m=16$ or $r/m=17$.  Two of these runs
  were compromised, so that this results in 98 configurations.  
\item {\bf Set $\mathbf{\cal S}_5$} contains 32 binaries with
  random spin-magnitudes $\chi_{A,B}\leq0.5$ at random spin-directions with
  random mass-ratios $q\in[1,2]$ for fixed $r/m=15$.  
The explicit parameters are listed in Table~\ref{tab:RunsRandom} in the Appendix.
\end{itemize}

Figure~\ref{fig:eccentricity-all} summarizes initial and final eccentricity of 
all configurations which were quasi-circularized. We initialized eccentricity 
reduction using an ad-hoc formula based on Kepler's law. As can be seen in 
Fig.~\ref{fig:eccentricity-all}, this reduced the starting eccentricity of 
${\cal S}_{1,2,3}$ by about a factor of 10 relative to that of ${\cal S}_0$.  
However, it did not significantly reduce the starting eccentricity of the 
precessing systems in ${\cal S}_{4,5}$.  A more comprehensive approach is 
therefore needed.

\section{Eccentricity reduction using fitting formulas}
\label{sec:fitting}

Kepler's law helped in quasicircularizing the binaries of the set ${\cal S}_0$ 
when small variations are introduced in either the separation or the mass ratio
of the binary. The Newtonian Kepler's law does not incorporate spin effects 
in the initial parameters of the binary. Therefore, we will base
fitting formulas for spinning binaries on post-Newtonian expansions 
for spinning binaries. 

\subsection{PN formulas}

\begin{figure}
\includegraphics[scale=0.51]{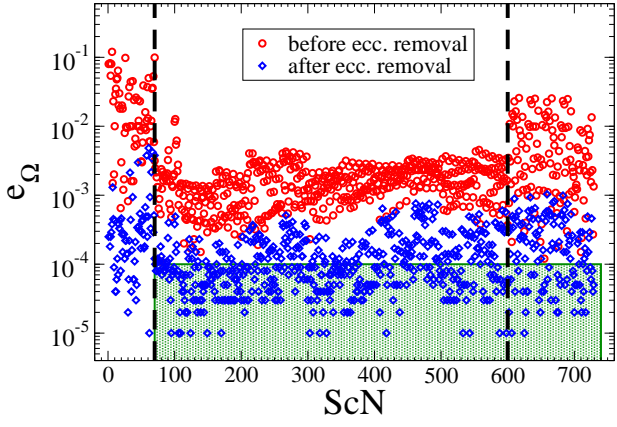}
\caption{\label{fig:eccentricity-all} 
{\bf Initial and final eccentricities of all runs in the paper.}
The first 69 configurations corresponds to the set ${\cal S}_0$ of binaries 
listed in Table~\ref{tab:Runs}. The next 530 configurations corresponds to
sets ${\cal S}_1, {\cal S}_2, {\cal S}_{3}$ of binaries, and the last 130 
configurations correspond to sets ${\cal S}_{4}$ and ${\cal S}_{5}$ with
random spin configurations.  The final eccentricity of the runs in sets 
${\cal S}_1,\ldots {\cal S}_{5}$ is reduced to less than $10^{-3}$.
}
\end{figure}

In the post-Newtonian approximations, the initial quasi-circular
parameters $\Omega_0$ or $\dot r_0$ (and $r_0$) are given as a series
expansion in $m/r$ (or in $\Omega_0$) for any physical configuration
for the binary.  We shall use the PN expressions derived by
Kidder~\cite{kidder95}.  In these expressions, nonspinning effects are
included up to 2nd post-Newtonian order, while the spin effects enter
at 1.5th and 2nd post-Newtonian order.  We define the unit vector
along the angular momentum ${\bf \hat L_N}= {\bf L_N/|L_N|}$ where
${\bf L_N} \equiv \mu ({\bf r}\times {\bf v})$ and the spin unit
vectors ${\bf \hat s}_{A,B}={\bf S}_{A/B} / S_{A/B}= {\bf \chi}_{A/B} /
\chi_{A/B}$.  Furthermore, let  $m=M_A+M_B$ be the total mass, $q=m_A/m_B$ the mass-ratio, $\mu= m_A m_B/m$ be the reduced mass and $\eta \equiv m_A m_B /m^2=q/(1+q)^2$ the symmetric mass-ratio.  
In terms of these quantities, Kidder~\cite{kidder95} derives
 the orbital frequency as a function of separation,

\begin{widetext}
\begin{eqnarray}
{\Omega^2 r^3 \over m} &=& 
1 
- \left(3-\eta \right) \left( {m \over r} \right) 
-\Bigg[  \sum_{i = A,B} \chi_i {\bf \hat L_N \cdot \hat s_i} \left( 2 {m_i^2 \over m^2} +3 \eta \right) \Bigg]
\left( {m \over r} \right)^{3/2} 
\nonumber \\&& 
+ \Biggl[ \left(6 + {41 \over 4}\eta + \eta^2 \right) 
- {3 \over 2}
\eta \chi_A \chi_B 
\Bigl( {\bf \hat s_A \cdot \hat s_B} - 3 {\bf \hat L_N \cdot \hat s_A} {\bf \hat L_N \cdot \hat s_B} \Bigr) 
\Biggr] \left( {m \over r} \right)^2 
\,.
\label{eq:inspiral-Omega}
\end{eqnarray}
The radial velocity as a function of separation is given by:
\begin{eqnarray} 
{-5  \over 64}  \eta^{-1} \left( {m \over r} \right)^{-3}  \dot r &=& 
1
- {1 \over 336} \big( 1751+588\eta \big) \left( {m \over r} \right)
- \Biggl\{  {7 \over 12} \sum_{i = A,B} \biggl[ 
\chi_i ({\bf \hat L_N \cdot \hat s_i}) ( 19{m_i^2 \over m^2} +15\eta) 
\biggr]  
- 4\pi \Biggr\} \left( {m \over r} \right)^{3/2} 
 \nonumber \\ && \mbox{}
- {5 \over 48}
\eta \chi_A \chi_B \left[ 59({\bf \hat s_A \cdot \hat s_B}) - 173({\bf \hat L_N
\cdot \hat s_A})({\bf \hat L_N \cdot \hat s_B}) \right]
\left( {m \over r} \right)^2\,. 
\label{eq:inspiral-rdot}
\end{eqnarray}
The orbital separation $r/m$ as a function of orbital frequency is 
\begin{eqnarray}
 (m\Omega)^{2/3}\frac{r}{m}  &=&
 1 - {1 \over 3} \left(3-\eta \right) (m\Omega)^{2/3}
- {1 \over 3} \sum_{i = A,B} \left[ \chi_i \left( 2{m_i^2 \over m^2} 
+ 3 \eta \right) {\bf \hat L_N \cdot \hat s_i}  \right]
m\Omega 
\nonumber \\ &&  
+ \left[
\eta  \left( {19 \over 4} + {1 \over 9}\eta \right)
- {1 \over 2}
\eta \chi_A \chi_B \left( {\bf \hat s_A \cdot \hat s_B} 
- 3 {\bf \hat L_N \cdot \hat s_A} {\bf \hat L_N \cdot \hat s_B} \right) \right]
(m\Omega)^{4/3}\,. 
\label{eq:inspiral-r}
\end{eqnarray}
And finally,  the number of orbits accumulated during inspiral from initial 
orbital frequency $\Omega_i$ to final frequency $\Omega_f$ is given by
$N = \left[\Psi(\Omega_i) -\Psi(\Omega_f)\right]/(2\pi)$
with orbital phase $\Psi$ satisfying
\begin{eqnarray}
32\eta\;\Psi(\Omega)  &=& 
 (m\Omega)^{-5/3}
+ {5 \over 1008}(743  + 924  \eta) 
(m\Omega)^{-1} 
%
+ \left\{ {5 \over 24} \sum_{i = A,B} \left[ \chi_i {\bf \hat L_N \cdot \hat s_i}
 \left( 113{m_i^2 \over m^2} +75\eta \right) \right] - 10\pi \right\} 
(m\Omega)^{-2/3} \nonumber \\ 
&& \qquad\qquad\quad
+ {5 \over 48} 
\eta \chi_A \chi_B \left( 247 {\bf \hat s_A \cdot \hat s_B} 
- 721 {\bf \hat L_N \cdot \hat s_A} \, {\bf \hat L_N \cdot \hat s_B} \right) 
(m\Omega)^{-1/3}.
\label{eq:numberoforbits} 
\end{eqnarray} 
\end{widetext} 

\subsection{NR Fitting Formulas }

The goal of this section is to provide fitting formulae that can
predict $\Omega_0$ and $\dot a_0$ as functions of symmetric mass-ratio
$\eta$, initial black hole spins ${\bf S}_{A/B}$ and initial black
hole distance $r/m$.  This is an 8-dimensional parameter space.  As we
shall see, linear fits are not sufficient, however, if one were to
write a straightforward 8-dimensional higher-order Taylor-series, the
number of coefficients would quickly explore with expansion order.
Therefore, one needs to be judicious to include only terms that
contribute to the fit.

Motivated by Eqs.~(\ref{eq:inspiral-Omega}) and~(\ref{eq:inspiral-rdot}), 
we do {\em not} write down fitting formulae directly for $\Omega_0$ and $\dot a_0$, but rather 
for the quantities 
\begin{equation}
\kappa_0 \equiv \frac{\Omega_0^2 r_0^3}{m}
\end{equation}
and
\begin{equation}
\rho_0 \equiv -\frac{5m}{64\eta} \left(\frac{r_0}{m}\right)^4 \dot a_0\,.
\end{equation}
For each quasi-circular configuration, we compute $\kappa_0$ and $\rho_0$.

Beginning to fit the non-spinning components, we only consider
  the quasi-circular configurations of set ${\cal S}_1$ 
and fit polynomials in $\eta$ and inverse distance $u\equiv m/r$:
\begin{align}
\label{eq:kappaNS}
\kappa_{\rm NS} &= \sum_{i,j=0}^{i+j\leq3} b_{i,j} \eta^i u^j\,,\\
\label{eq:rhoNS}
\rho_{\rm NS} & = \sum_{i,j=0}^{i+j\leq3} a_{i,j} \eta^i u^j\,.
\end{align}
(Here, the subscript 'NS' refers to non-spinning binaries).
  We found that the triangular truncation $i+j\le 3$ gave a good fit
  with a reasonably low number of coefficients.  The coefficients $a_{i,j}$ and $b_{i,j}$ are listed in Tables~\ref{tab:aij} and~\ref{tab:bij}.


For spinning binaries of sets ${\cal S}_2$, ${\cal
  S}_3$, ${\cal S}_4$ and ${\cal S}_5$, we first subtract the fits
Eqs.~(\ref{eq:kappaNS}) and~(\ref{eq:rhoNS}), i.e. we compute for each spinnning quasi-circular configuration
\begin{align}
\delta \kappa_S &\equiv \kappa_0 -\kappa_{NS}|_{\eta_0,u_0}\,,\\
\delta \rho_S &\equiv \rho_0 - \rho_{NS}|_{\eta_0,u_0}\,.
\end{align}

The functions $\delta\kappa_S$ and $\delta\rho_S$ should be functions
of the spins of the black holes, mass-ratio and separation.  These
functions should vanish for zero spins, to reduce to the fit for non-spinning BBH.

We will use in total 14 basis-functions to represent the
  spin-sector, which we shall label $\hat e_\alpha$, $\alpha=1,
  \ldots, 14$.  The first four basis-functions take the functional form of the spin-terms in Eqs.~(\ref{eq:inspiral-Omega})--(\ref{eq:numberoforbits}):
\begin{subequations}
\begin{align}
\hat e_1&= {\bf S}_A\cdot{\bf \hat L},\\
\hat e_2&= {\bf S}_B\cdot{\bf \hat L},\\
\hat e_3&= {\bf S}_A\cdot{\bf S}_B,\\
\hat e_4&= {\bf S}_{A,\perp}\cdot{\bf S}_{B,\perp},
\end{align}
where ${\bf S}_{i,\perp}={\bf S}_i - {\bf \hat L}({\bf\hat L}\cdot{\bf S}_i), i=A,B$ are the 
projections of each black hole spin into the orbital plane.

The next two basis-functions account for the the angle between ${\bf
  S}_{A/B}$ relative to the other black hole:
\begin{align}
\hat e_5&= {\bf S}_A\cdot{\bf \hat r},\\
\hat e_6&= {\bf S}_B\cdot{\bf \hat r},
\end{align}
where ${\bf \hat r}$ is the unit-vector pointing from black hole 2 to
black hole 1. These
terms are not present in the post-Newtonian expansion.  However, we
find them to be necessary for a good fit.  We
conjecture that the origin of these terms rests in details of how
numerical BBH initial data is constructed.  For instance,
conformally flat quasi-equilibrium BBH initial data requires a
somewhat different choice of $\dot a_0$ to achieve low eccentricity
compared to superposed Kerr-Schild
data~\cite{Garcia:2012dc,Lovelace2008}.  Both types of initial-data
require some offset in $\dot a_0$ relative to post-Newtonian
expansions, and the need to include $\hat e_5$ and $\hat e_6$ when
fitting for low-eccentricity initial data parameters indicates that
this offset depends on the angle between black hole spin and direction
to the partner black hole.

Finally, as demonstrated below, the fit of the spin-sector can be
improved by including certain higher-order basis-functions:
\begin{align}
\label{eq:hate_7}
\hat e_7&= ({\bf S}_A\cdot{\bf \hat L})^2,\\
\hat e_8&= ({\bf S}_A\cdot{\bf \hat L})^3,\\
\hat e_9&= ({\bf S}_B\cdot{\bf \hat L})^2,\\
\hat e_{10}&= ({\bf S}_B\cdot{\bf \hat L})^3,\\
\hat e_{11}&= ({\bf S}_A\cdot{\bf S}_B)^2,\\
\hat e_{12}&= ({\bf S}_{A,\perp}\cdot{\bf S}_{B,\perp})^2,\\
\hat e_{13}&= ({\bf S}_A\cdot{\bf \hat r})^2,\\
\label{eq:hate_14}
\hat e_{14}&= ({\bf S}_B\cdot{\bf \hat r})^2.
\end{align}
\end{subequations}

The functions $\delta \kappa_S$ and $\delta \rho_S$ are expanded into 
a series that depends on $\eta$, $u$ and the spin basis $\hat e_\alpha$:
\begin{align}
\label{eq:deltakappaS-fit}
\delta \kappa_{\rm S} &= \sum_{i=0}^3\sum_{j=2}^5\sum_{\alpha=1}^{14} c_{i,j,\alpha} \eta^i u^{j/2} {\hat e}_\alpha,\\
\label{eq:deltarhoS-fit}
\delta \rho_{\rm S} &= \sum_{i=0}^3\sum_{j=2}^5\sum_{\alpha=1}^{14} d_{i,j,\alpha} \eta^i u^{j/2} {\hat e}_\alpha.
\end{align}
These expansions have 224 coefficients each.  For our best result, we
fit these coefficients against that subset of configurations of the sets ${\cal S}_{3,4,5}$ which have $e_\Omega<10^{-4}$.  These configurations are located in the green shaded region of Fig.~\ref{fig:eccentricity-all}.  The
resulting coefficients are listed in
Tables~\ref{tab:cijk} and~\ref{tab:dijk} in the Appendix.

\subsection{Quality of the fitting formulas}

\begin{figure}
\includegraphics[scale=0.52]{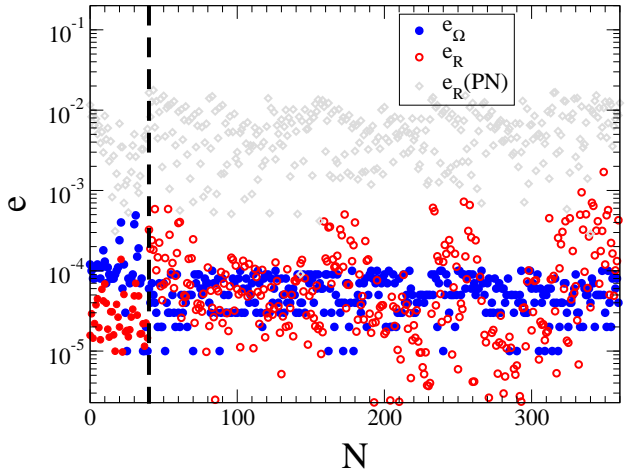}
\caption{
\label{fig:eccentricity-tests-eR-e0-ePN} 
Eccentricities $e_\Omega$ and $e_R$ for the binaries of the sets
${\cal S}_2$, and those configurations of ${\cal S}_{3,4,5}$ with
$e_\Omega\leq 0.0001$.  The residuals $e_R$ of the fitting formulas
are comparable to the actual estimated eccentricity
$e_\Omega$. Further, improvement of the fitting formulas requires
quasi-circular configuration with lower eccentricities $e_\Omega$.
The first 40 simulations are nonspinning. The expected PN
eccentricities are plotted in grey.  }
\end{figure}

The quasi-circular orbital parameters that form the basis of
  the fits Eqs.~(\ref{eq:deltakappaS-fit})
  and~(\ref{eq:deltarhoS-fit}) are not perfectly accurate; rather they
  contains residual eccentricity as indicated in
  Fig.~\ref{fig:eccentricity-all}.  Therefore, we should expect the
  low-eccentricity fits to be good to only a comparable level of
  precision.  If the fits were more accurate than indicated by
  Fig.~\ref{fig:eccentricity-all}, this would indicate a fit to the
  {\em errors} in the numerical data, which would not carry any
  additional physical information.  To quantify the quality of the fits, 
  we translate the residuals 
\begin{align}
  \Delta\Omega_0 &= \Omega_{0,\rm NR}-\Omega_{0,\rm fit},\\
  \Delta\dot a_0 &= \dot a_{0,\rm NR}-\dot a_{0,\rm fit}
\end{align}
into an equivalent eccentricity via Eq.~(\ref{eq:new-epsNewt}):
\begin{equation}
e_R= \left[ \left(\frac{\Delta\dot a_0}{\Omega_{0,\rm NR}}\right)^2 
  +\left(2\frac{\Delta\Omega_0}{\Omega_{0,\rm NR}}\right)^2 
  \right]^{1/2}.
\end{equation}
Our target is that $e_R$ as computed from the fits is roughly
comparable to the eccentricity $e_\Omega$ of the data used to compute
the fits.  If $e_R\gg e_\Omega$ the fit is not as accurate as it could be; if $e_R\ll e_\Omega$, the fit has so many degrees of freedom that it fits the residual eccentricity 
of the NR simulations.  Figure~\ref{fig:eccentricity-tests-eR-e0-ePN}
demonstrates that indeed $e_{\rm R} \sim e_\Omega$, as desired.

\begin{figure}
\includegraphics[scale=0.47]{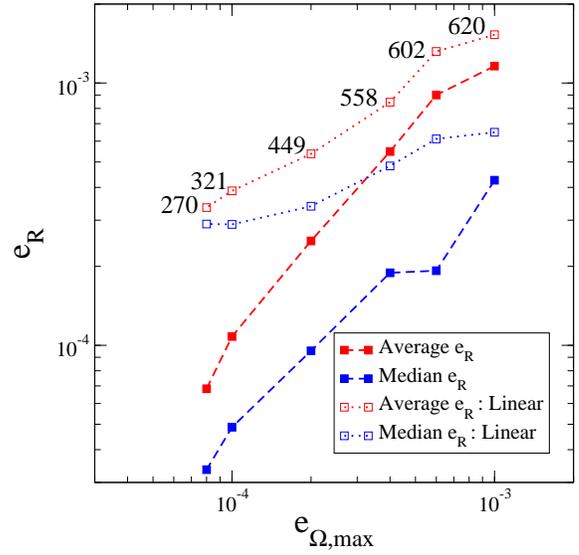}
\caption{\label{fig:eccentricity-tests-convergence} 
{\bf Performance of fit vs. the eccentricity of the configurations used in the fit.}  
Eccentricity fits are performed only using those
configurations with $e_\Omega< e_{\Omega,\rm max}$.  As a function of 
$e_{\Omega,\rm max}$ the plot shows the average and the median of $e_R$. 
Reducing the eccentricity $e_{\Omega,\rm max}$ improves the fitting formulas 
performance. The numbers indicate the number of configurations used in each fit.
}
\end{figure}

The dependence of the quality of the fit onto the
  configurations used for the fit is further explored in
  Fig.~\ref{fig:eccentricity-tests-convergence}.  For this figure, we
  perform the fit for the spinning sector $\delta\kappa_S$,
  $\delta\rho_S$ several times, based on those configurations within
  ${\cal S}_{3,4,5}$ with eccentricity $e_\Omega\le e_{\Omega,\rm
    max}$.  The dashed curves of
  Fig.~\ref{fig:eccentricity-tests-convergence} show the residual
  eccentricity $e_R$ of these fits as a function of threshold
  eccentricity $e_{\Omega,\rm max}$.  For the dashed lines the
  residual of the fit $e_R$ is similar to $e_{\Omega,\rm max}$,
  indicating that the fit works as well as can be expected.  If the
  basis-functions for the spin-sector are restricted to only the
  linear terms, $\hat e_\alpha, \alpha=1,\ldots, 6$, then we obtain
  the dotted lines.  Now the residual eccentricity is several times
  larger than $e_{\Omega,\rm max}$, indicating the possibility to
  improve the fit. 

\begin{figure}
\includegraphics[scale=0.47]{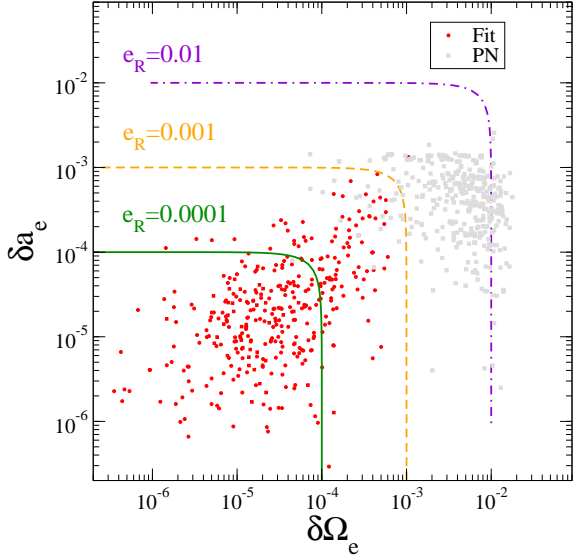}
\caption{
\label{fig:eccentricity-tests} 
{\bf Eccentricities $(\delta a_e,\delta\Omega_e)$ for the binaries
of the sets ${\cal S}_1, \cdots, {\cal S}_5$ with $e\leq 0.0001$} as 
predicted by the 
the fitting formulas and the PN expressions. Most binaries have an
eccentricity less than 0.0001 when using the fitting formulas, while
PN expressions generate binaries with nearly 0.01 eccentricity.
The fits yield lower eccentricity than PN expansions, with the
improvement primarily due to the more accurate prediction
of radial velocity ($\delta a_e$). 
We plot the fits based on binaries with eccentricity less than 0.0001.
}
\end{figure}

We now compare our low-eccentricity fits with post-Newtonian
  expansions that predict initial data parameters.  Using
  Eqns.~(\ref{eq:inspiral-Omega}) and~(\ref{eq:inspiral-rdot}) we
  compute $\Omega_{\rm PN}$ and $\dot a_{\rm PN}$ for each
  configuration.  We then compute the difference between eccentricity-reduced NR parameters and the PN parameters,
\begin{align}
\label{eq:delta-Omega-e-PN}
\delta\Omega_e&\equiv  2\frac{\Omega_{0,\rm PN}-\Omega_{0,\rm NR}}{\Omega_{0,\rm NR}},\\
\label{eq:delta-adot-e-PN}
\delta\dot a_e&\equiv  \frac{\dot a_{0,\rm PN}-\dot a_{0,\rm NR}}{\Omega_{0,\rm NR}},\\
\label{eq:e_RPN}
e_{R,\rm PN}&\equiv \sqrt{ \delta\Omega_e^2 + \delta\dot a_e^2}.
\end{align}
Because we know low-eccentricity orbital parameters $\Omega_{0,\rm
  NR}$, $\dot a_{0,\rm NR}$ these formulae allow us to estimate the
eccentricity the PN parameters would have {\em without} having to
evolve with the PN initial-data parameters.  
Figure~\ref{fig:eccentricity-tests-eR-e0-ePN} shows this estimated
eccentricity $e_{R,\rm PN}$ as the grey dots.  Our fits improve the
post-Newtonian initial data parameters by about two orders of
magnitude, consistently for all spin-directions, including precessing
binaries.

Figure~\ref{fig:eccentricity-tests} plots separately the two
components $\delta\Omega_e$ and $\delta\dot a_e$,
cf. Eqs.~(\ref{eq:delta-Omega-e-PN})
and~(\ref{eq:delta-adot-e-PN}). The plot shows the PN data in grey and
the results of the fitting formulas as red circles.  Once again, the
improvement gained by the fits is apparent: The eccentricity is
smaller by nearly two order of magnitude relative to PN initial-data
parameters.  We also note that for the fit, $\delta \dot a_e$ and
$\delta\Omega_e$ have approximately equal magnitudes.   For PN, in contrast,
$\delta\Omega_e$ is about a factor $\sim 10$ larger than $\delta\dot a_e$,
indicating that an error in the orbital frequency is the major cause
of eccentricity when using PN parameters. 

\subsection{Number of orbits fits}

 When selecting initial data parameters for an evolution, it
  is useful to be able to estimate of how long the simulation will be.
  One widely used measure is the number of orbits the binary completes
  before merger.  We have evolved the binaries in sets ${\cal S}_0$ and ${\cal S}_5$ through inspiral phase, so we can use this information to 
  derive a fitting formula for the number of orbits.  We base this fitting formula on Eqs.~(\ref{eq:numberoforbits}), specifically
\begin{equation}\label{eq:Norbits-fit}
N=\frac{1}{2\pi}\left(\Psi_{\rm fit}(\Omega_i) - \Psi_{\rm fit}(\Omega_f)\right).
\end{equation}
The fitting parameters $d_1,\ldots d_9$ are incorporated into the phase-formula, 
\begin{widetext}
\begin{align}
32\eta\,\Psi_{\rm fit}(\Omega)  =& (m\Omega)^{-5/3} 
\!+ \frac{5(743 d_1 + 924 d_2 \eta)}{1008} (m\Omega)^{-1}
\!+\! \left\{ \sum_{i} \left[ \frac{5\chi_i}{24} {\bf \hat L_N \!\cdot\! \hat s_i}
 \Big( 113d_7{m_i^2 \over m^2} +75d_8\eta \Big) \right] \!-\! 10\pi d_3\! \right\} 
(m\Omega)^{-2/3}  \nonumber \\ 
&+ \left[(d_4+d_5\eta+d_6\eta^2)
+ {5 \over 48} d_9
\,\eta\, \chi_A\, \chi_B \left( 247 {\bf \hat s_A \!\cdot\! \hat s_B} 
- 721 {\bf \hat L_N \!\cdot\! \hat s_A} \, {\bf \hat L_N \!\cdot \!\hat s_B} \right) 
\right]%
(m\Omega)^{-1/3}.
\label{eq:Phase-fit} 
\end{align} 
\end{widetext}

The number of orbits is fitted to Eqs.~(\ref{eq:Norbits-fit})
  and~(\ref{eq:Phase-fit}) using the inspiral data from sets ${\cal
    S}_0$ and ${\cal S}_{5}$ as given in Tables~\ref{tab:Runs}
  and~\ref{tab:RunsRandom}.  We include only those inspirals that
  reach orbital frequency $0.05/m$ or greater, that inspiral for at
  least 10 orbits and that have an eccentricity of less than 0.005.

\begin{figure}
\includegraphics[scale=0.51]{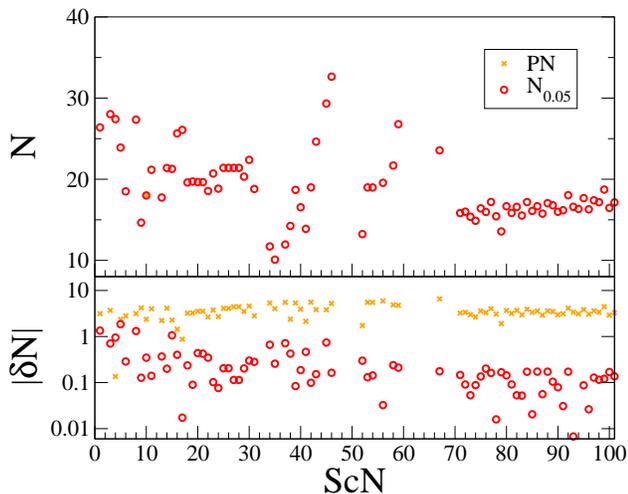}
\caption{\label{fig:number-of-orbits} 
{\bf The number of orbits of the runs and the residual of the fitting
formulas.} In the upper panel, we plot the number of orbits $N_{0.05}$
through the inspiral in red between the initial orbital frequency $\Omega_i$
and the final orbital frequency $\Omega=0.05/m$ evolved. 
In the lower panel, we plot the difference between the fitting polynomial 
and the numerical data of the number of orbits in red. We compare the results 
to the PN approximation ploted in orange.
}
\end{figure}

The results of the fit for $N$ are shown in Fig.~\ref{fig:number-of-orbits}.
The upper panel of Fig.~\ref{fig:number-of-orbits} 
plot the number of orbits evolved between the initial 
orbital frequency $\Omega_i$ and the final orbital frequency $0.05/m$ 
for the inspirals used in these fits. 
For the first 69 simulations of set ${\cal S}_0$, binaries with large 
differences in the initial separation and mass ratios are evolved. Therefore, 
a large variation in the number of orbits is observed. For the last 32 runs of 
set ${\cal S}_5$, all binaries start with the same initial separation 
but with a random mass ratio and spin orientation and magnitude. In this case, 
their number of orbits varies between 15 and 20.

In the lower panel of Fig.~\ref{fig:number-of-orbits}, we plot the difference 
between the number of orbits measured up to a certain orbital frequency, and
the number predicted by the fitting formula~(\ref{eq:Norbits-fit}) as red 
points. This residual is less than 1 orbit for the set 
${\cal S}_0$, and it is less than 0.2 orbits for the binaries of set 
${\cal S}_{5}$.  No variation in the quality of the fits is noticed as 
we changed the final frequency from $0.05/m$.
We use also the PN expression in Eq.~\ref{eq:numberoforbits} to plot the 
difference in the number of orbits between the numerical simulations and the PN
predictions measured at an orbital frequency of $0.05/m$ in orange. The initial
frequency used is the value for which the simulation starts with, i.e. 
$\Omega_i$.  The straight post-Newtonian formula predicts a number of orbits
that generally differs by 3 to 6 orbits from the NR result.

\section{Generating initial data with a pre-determined eccentricity}
\label{sec:eccentricity} 

\begin{figure}
\includegraphics[scale=0.51]{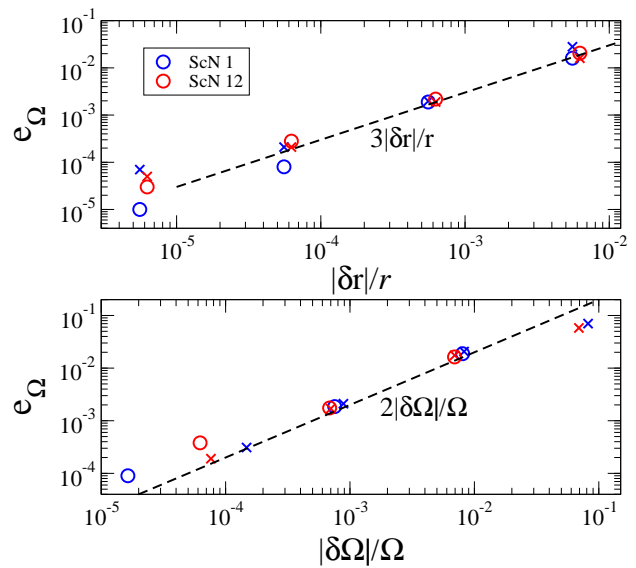}
\caption{\label{fig:Set-ecc} Eccentricity of runs with initial data parameters perturbed from their low-eccentricity values.  Two representative configurations are shown, ScN 1 (equal mass, zero spin), and ScN 12 (q=1.5, precessing spin).
Circles are for positive variations, while crosses are for negative variation.
The dashed line represents the expectation, Eq.~(\ref{eq:new-eNewt}).
}
\end{figure}

So far, we have been concerned with achieving BH-BH
  simulations with very small eccentricity.  Let us now consider how
  to choose initial-data parameters that result in some desired {\em
    non-zero} eccentricity.  If one starts from known
  zero-eccentricity orbital parameters $(r_0, \Omega_0, \dot a_0)$,
  and perturbs these, $(r_0+\delta r, \Omega_0+\delta\Omega, \dot
  a_0+\delta\dot a)$, then the Newtonian formula
  Eq.~(\ref{eq:new-eNewt}) should give a reasonable approximation of
  the resulting eccentricity.  To test this assumption we use two
  eccentricity-removed configurations, ScN 1 (equal mass, zero spin)
  and ScN 12 (mass-ratio 1.5, precessing).  We perturb first by
  changing the initial separation while keeping $\Omega_0$ and $\dot
  a_0$ constant, and, second, by changing $\Omega_0$ while keeping
  $d_0$ and $\dot a_0$ constant.  For each perturbation, we solve for
  a new initial-data set, and evolve long enough to measure
  $e_\Omega$.  The results are plotted in Fig.~\ref{fig:Set-ecc},
  along with the result of Eq.~(\ref{eq:new-eNewt}), $e=3|\delta r|/r$
  and $e=2|\delta\Omega|/\Omega$, respectively.

The agreement between the full numerical simulation and the
  Newtonian formula is very good.  The exceptions are very low
  eccentricity $e_\Omega\lesssim 10^{-4}$, where measuring
  eccentricity in the numerical simulation is difficult, and for very
  large eccentricity $e\sim 0.1$, where linear perturbations may no
  longer be adequate.  
Besides controlling eccentricity, one can also use
  Eq.~(\ref{eq:new-phi0Newt}) to control the phase or periastron. 

  So far, we have perturbed off a zero-eccentricity configuration.  As
  we have seen in Sec.~\ref{sec:ecc_NR}, our fitting formulae for
  initial data-parameters result in eccentricities $e\lesssim
  10^{-3}$.  If one is interested in eccentricities larger than this
  value, then one can also perturb around the results of the
  low-eccentricity fitting formulas. This now allows to obtain a BH-BH
  simulation with desired length (via Eq.~\ref{eq:numberoforbits}) and
  desired eccentricity, via the fitting formulas for low-eccentricity
  parameters and Eq.~(\ref{eq:new-eNewt}) and~(\ref{eq:new-phi0Newt}),
  {\em without} any iterative procedures.

\section{Conclusions}
\label{sec:conclusions}

 Evolutions of binary black holes start from initial data.
  Choices that enter the initial data determine the orbital
  eccentricity of the subsequent evolution, and the length of the
  inspiral (i.e. the number of orbits to merger).  This paper presents
  techniques that allow to choose initial data parameters that result
  in very small eccentricity ($e\sim 10^{-4}$) with an inspiral of a
  desired number of orbits.  We also present techniques that allow to
  choose initial data parameters that result in a desired non-zero
  eccentricity.  Our techniques are applicable for {\em generic}
  precessing binary black holes of mass-ratios $q\lesssim 8$ and for
  spin-magnitudes $\lesssim 0.5$.  Because of fitting formulae we
  develop here, the use of these techniques requires no significant
  computational cost.

 This paper presents quasi-circular initial data for 729
  different configurations (mass-ratio, separation, spins) of binary
  black holes.  This study covers for the first time a full
  7-dimensional {\em voluminous} region of parameter space of generic
  non-eccentric BBH inspirals, rather than just lower-dimensional
  subspaces like aligned spin binaries (e.g.~\cite{Ajith:2012tt}).  
The mass ratio varies between
  1 and 8 for these simulations, 620 binaries are spinning, most of
  these with generic, precessing spins of dimensionless magnitude up
  to 0.5.  

 The orbital eccentricity of all 729 configurations is
  iteratively reduced using techniques developed in earlier
  work~\cite{Pfeiffer-Brown-etal:2007,Boyle2007,Buonanno:2010yk}.  To
  remove or ease the computational burden of iterative eccentricity
  removal, we introduce fitting formulas that predict low-eccentricity
  orbital parameters.  
  Fitting the 2-dimensional non-spinning sector spanning mass-ratio
  and spin requires 10 fitting parameters each in the formulae for
  orbital frequency and radial velocity.  Extending this fit to the
  8-dimensional spinning-sector (mass-ratio, separation, two spin
  vectors) required 224 fitting parameters each.  Perhaps
  surprisingly, despite the comparatively low spins considered here
  $\chi\lesssim 0.5$, higher order corrections to the spin
  incorporated via Eqs.~(\ref{eq:hate_7})--(\ref{eq:hate_14})
  noticeably improve the quality of the fit,
  cf. Fig.~\ref{fig:eccentricity-tests-convergence}.
We also provide a fitting formula for the expected number of orbits
during the inspiral of low-eccentricity initial data, again for
generic precessing spins and mass-ratios.

The fitting formulas allow to achieve BH-BH configurations of
  desired initial separation (or desired initial orbital frequency or
  desired number of orbits) with an eccentricity of $\sim 10^{-4}$
  without having to perform any iterative runs.  If
lower eccentricity is desired, it is always possible to refine by
iterative eccentricity removal.

Based on these fits, we develop a technique to predict the
  eccentricity of initial-data parameters without having to evolve the
  initial data at all.  This technique is based on the deviation of
  the initial-data parameters from those for low-eccentricity (the latter determined either from our bank of 729 configurations, or computed from our low-eccentricity fits),
  cf. Eqs.~(\ref{eq:delta-Omega-e-PN})--(\ref{eq:e_RPN}).  It allows
  us to estimate the orbital eccentricity that would have resulted
  from post-Newtonian initial data parameters without constructing
  initial data for those configurations.  The results are shown in
  Figs.~\ref{fig:eccentricity-tests-eR-e0-ePN}
  and~\ref{fig:eccentricity-tests} demonstrating that our fitting formulas
  yield orbital eccentricity about two orders of magnitude smaller
  than post-Newtonian expansions.  

The ability to estimate eccentricity without an evolution is also useful
in another scenario: To obtain an evolution with desired eccentricity, 
one can simply perturb the low-eccentricity fits and use Eqs.~(\ref{eq:delta-Omega-e-PN})--(\ref{eq:e_RPN}) to determine the eccentricity of the perturbed initial-data.  This procedure allows to construct initial data of desired eccentricity without having to perform any intermediate evolutions.  This procedure is demonstrated in Sec.~\ref{sec:eccentricity}.

During the construction of BBH initial-data, free parameters like
initial orbital frequency and initial radial velocity must be chosen.
Early in the evolution, the space-time relaxes to a quasi-stationary
steady state, changing the black hole parameters slightly and emitting
a pulse of ``junk radiation''~\cite{Lovelace2009}.  This relaxation
causes orbital frequency and radial velocity (and also black hole
masses and spins, albeit to a smaller degree~\cite{Chu2009}) to
deviate from the corresponding parameters specified in the
initial-data.  These drifts depend on the precise type of initial data
evolved (e.g. conformally flat conformal-thin sandwich data, as here;
super-posed Kerr-Schild data; puncture data) and may exhibit a
dependence on black hole parameters {\em different} from the usual
post-Newtonian terms.  These drifts in parameters, induced by specific
choices of the initial-data construction, may very well be the reason
why our fitting formulas require terms that are not present in
post-Newtonian expansions, cf. Eqs.~(\ref{eq:kappaNS})
and~(\ref{eq:rhoNS}).  Because these drifts ---as well as the
coordinate systems used when constructing initial data--- are
different for each class of initial-data, we expect that a similar
fitting effort is necessary for superposed Kerr-Schild data, which
allows access to higher black-hole spins.

Out of the 729 configurations, we evolve 101 simulations fully through
inspiral.  Only the number of orbits of each inspiral was used here in
order to write fitting formulas for any spinning configurations.
Because of the wealth of information in these simulations, they will
form the basis of a large number of future investigations into
periastron-advance, black hole remnant properties, analytic template
modelling, and gravitational wave data-analysis efforts.

\begin{acknowledgments}
We would like to thank Mike Boyle, Tony Chu, Larry Kidder, Geoffrey Lovelace, 
Mark Scheel, Bela Szilagyi and Nick Taylor for useful discussions. We 
gratefully acknowledge support from the NSERC of Canada, from the Canada 
Research Chairs Program, and from the Canadian Institute for Advanced Research.
Computations were performed on the GPC supercomputer at the SciNet HPC 
Consortium. SciNet is funded by: the Canada Foundation for Innovation under 
the auspices of Compute Canada; the Government of Ontario; Ontario Research 
Fund - Research Excellence; and the University of Toronto.
Results obtained in this paper were produced using the Spectral Einstein Code 
(SpEC)~\cite{SpECwebsite}.
\end{acknowledgments}

\appendix

\begin{widetext}
\section*{Appendix: Tables}

\setlength{\LTcapwidth}{0.9\textwidth}
\begin{longtable}[c]{|c|c|c|ccc|ccc|c|c|c|c|c|c|}
\caption{\label{tab:Runs}  
  {\bf Parameters of the runs evolved in ${\cal S}_0$. }
  The first column is the label for each configuration (we refer to individual runs as ``Sc21''), 
  the following columns show the mass ratio 
  $q$, the separation $r/m$, the initial orbital frequency $\Omega_i$, the 
  dimensionless expansion factor $\dot a$, the spin components of the first hole
  $\chi_A$ and 
  the spin of the second hole $\chi_B$, the eccentricity $e_\Omega$, the final 
  orbital frequency $\Omega_f$, the number of orbits $N_f$ between the initial
  frequency $\Omega_i$ and the final frequency $\Omega_f$.  
  The last column denotes the number of orbits between the initial orbital frequency until the 
  orbital frequency $0.05/m$.
 }\\
\hline
ScN & $q$ & $r/m$ & $\chi_A$ & $\theta_A/\pi$ & $\phi_A/\pi$ & $\chi_B$ & $\theta_B/\pi$ & $\phi_B/\pi$ & $10^2m\Omega_i$ & $10^{5}\dot a$ & $10^{4}e_\Omega$ & $10^2m\Omega_f$ & $N_f$ & $N_{0.05}$  \\ \hline
\endfirsthead
\multicolumn{15}{l}%
	{\tablename\ \thetable\ -- \textit{Continued}} \\
	\hline
ScN & $q$ & $r/m$ & $\chi_A$ & $\theta_A/\pi$ & $\phi_A/\pi$ & $\chi_B$ & $\theta_B/\pi$ & $\phi_B/\pi$ & $10^2m\Omega_i$ & $10^{5}\dot a$ & $10^{4}e_\Omega$ & $10^2m\Omega_f$ & $N_f$ & $N_{0.05}$  \\ \hline
\endhead
\hline \multicolumn{15}{r}{\textit{Continued}}\\ 
\endfoot
\endlastfoot
1 &  1  &  18  & 0 & 0 & 0 & 0 & 0 & 0 & 	 1.2202  &  -2.54  &  2.5  &  12.  &  27.9    &  26.38   \\  
2 &  1  &  19  & 0 & 0 & 0 & 0 & 0 & 0 & 	 1.1292  &  -2.10  &  1.8  &  3.3  &  28.4    &  -   \\  
3 &  1  &  18  & 0.5 & 0 & 0 & 0 & 0 & 0 & 	 1.2168  &  -1.63  &  2.5  &  7.5  &  29.2    &  28.02   \\  
4 &  1  &  19  & 0.5 & 0.5 & 0 & 0 & 0 & 0 & 	 1.1283  &  -1.88  &  2.9  &  8.0  &  31.7    &  27.41   \\  
5 &  1  &  19  & 0.5 & 1 & 0 & 0 & 0 & 0 & 	 1.1313  &  -2.80  &  3.8  &  4.2  &  28.1    &  23.90   \\  
6 & 1.345&16&0.320&0.667&0.142&0.150&0.732&0.148&1.4463  &  -3.60  &  2.4  &  12.  &  19.94  &  18.50   \\  
7 &  1.5  &  16  & 0 & 0 & 0 & 0 & 0 & 0 & 	 1.4427  &  -3.64  &  13.0 &  13.  &  21.1    &  19.50  \\  
8 &  1.5  &  18  & 0 & 0 & 0 & 0 & 0 & 0 & 	 1.2199  &  -2.15  &  4.5  &  13.  &  28.9    &  27.34  \\  
9 &  1.5  &  14  & 0.5 & 0 & 0 & 0 & 0 & 0 & 	 1.7367  &  -5.14  &  0.9  &  6.3  &  15.5  &  14.65   \\  
10 & 1.5  &  16  & 0.5 & 0.833 & 0 & 0 & 0 & 0 & 	 1.4481  &  -5.11  &  4.3  &  7.5391  &  18.8    &  18.00  \\  
11 & 1.5  &  16  & 0.5 & 0.167 & 0 & 0 & 0 & 0 & 	 1.4384  &  -2.94  &  0.4  &  7.6682  &  22.4    &  21.16   \\  
12 & 1.5  &  16  & 0.5 & 0.5 & 0 & 0 & 0 & 0 & 	 1.4427  &  -3.64  &  0.4  &  9.8  &  20.9    &  -   \\  
13 & 1.5  &  16  & 0.5 & 1 & 0 & 0 & 0 & 0 & 	 1.4490  &  -4.76  &  0.6  &  7.5  &  18.5    &  17.74   \\  
14 & 1.5  &  16  & 0.5 & 0 & 0 & 0 & 0 & 0 & 	 1.4379  &  -2.72  &  1.2  &  7.1  &  22.5    &  21.38   \\  
15 & 1.5  &  17  & 0.5 & 1 & 0 & 0 & 0 & 0 & 	 1.3286  &  -3.82  &  0.7  &  7.7  &  22.0    &  21.27   \\  
16 & 1.5  &  18  & 0.5 & 0.5 & 0 & 0 & 0 & 0 & 	 1.2192  &  -2.11  &  3.3  &  9.4  &  28.9    &  25.65   \\  
17 & 1.5  &  19  & 0.5 & 1 & 0 & 0 & 0 & 0 & 	 1.1321  &  -2.26  &  4.3  &  7.6  &  30.3    &  26.08   \\  
18 & 1.5  &  16  & 0.5 & 0.5 & 0 & 0.5 & 0.5 & 1 & 	 1.4424  &  -3.31  &  0.8  &  5.6  &  20.0    &  19.60   \\  
19 & 1.5  &  16  & 0.5 & 0.5 & 0 & 0.5 & 0.5 & 0 & 	 1.4414  &  -3.34  &  4.6  &  5.5  &  20.1    &  19.72   \\  
20 & 1.5  &  16  & 0.5 & 0.5 & 0 & 0.5 & 0.5 & -0.5 & 	 1.4534  &  -3.60  &  2.7  &  6.8  &  20.5    &  19.65   \\  
21 & 1.5  &  16  & 0.5 & 0.5 & 0 & 0.5 & 0.5 & 0.5 & 	 1.4533  &  -3.61  &  1.5  &  7.8  &  20.7    &  19.64   \\  
22 & 1.5  &  16  & 0.5 & 0.5 & 0 & 0.5 & 1 & 0 & 	 1.4455  &  -5.02  &  2.8  &  5.7  &  18.9    &  18.54   \\  
23 & 1.5  &  16  & 0.5 & 0.5 & 0 & 0.5 & 0 & 0 & 	 1.4388  &  -2.62  &  2.7  &  6.6  &  21.6    &  20.70   \\  
24 & 1.5  &  16  & 0.5 & 1 & 0 & 0.5 & 0 & 0 & 	 1.4446  &  -3.48  &  1.2  &  7.3  &  19.7    &  18.84  \\  
25 & 1.5  &  16  & 0.5 & 0 & 0 & 0.5 & 0.5 & 1 & 	 1.4370  &  -2.74  &  2.2  &  6.8  &  22.4    &  21.40  \\  
26 & 1.5  &  16  & 0.5 & 0 & 0 & 0.5 & 0.5 & 0 & 	 1.4370  &  -2.74  &  2.2  &  6.8  &  22.4    &  21.40  \\  
27 & 1.5  &  16  & 0.5 & 0 & 0 & 0.5 & 0.5 & -0.5 & 	 1.4485  &  -2.57  &  1.4  &  5.5  &  21.8    &  21.40  \\  
28 & 1.5  &  16  & 0.5 & 0 & 0 & 0.5 & 0.5 & 0.5 & 	 1.4485  &  -2.57  &  1.4  &  5.5  &  21.8    &  21.40  \\  
29 & 1.5  &  16  & 0.5 & 0 & 0 & 0.5 & 1 & 0 & 	 1.4402  &  -3.93  &  0.2  &  6.8  &  21.2    &  20.32  \\  
30 & 1.5  &  16  & 0.5 & 0 & 0 & 0.5 & 0 & 0 & 	 1.4344  &  -1.56  &  0.6  &  7.1  &  23.72   &  22.38  \\  
31 & 1.5  &  16  & 0.320 & 0.667 & 0.142 & 0.150 & 0.732 & 0.148 &  1.4463  &  -3.60  &  2.2  &  11.  &  20.2   &  18.80  \\  
32 & 3  &  14  & 0 & 0 & 0 & 0 & 0 & 0 & 	 1.7427  &  -4.69  &  21.2  &  8.3  &  17.6    &  16.36  \\  
33 & 3  &  12  & 0.5 & 0.833 & 0 & 0 & 0 & 0 & 	 2.1817  &  -11.5  &  4.4  &  8.3  &  9.3    &  8.36  \\  
34 & 3  &  12  & 0.5 & 0.167 & 0 & 0 & 0 & 0 & 	 2.1522  &  -7.72  &  0.4  &  9.2  &  13.7    &  11.71  \\  
35 & 3  &  12  & 0.5 & 0.5 & 0 & 0 & 0 & 0 & 	 2.1644  &  -8.89  &  1.6  &  7.9  &  11.4    &  10.09  \\  
36 & 3  &  12  & 0.5 & 1 & 0 & 0 & 0 & 0 & 	 2.1846  &  -12.3  &  9.4  &  10.  &  9.1    &  8.08   \\  
37 & 3  &  12  & 0.5 & 0 & 0 & 0 & 0 & 0 & 	 2.1508  &  -7.74  &  0.6  &  9.3  &  14.1    &  11.95  \\  
38 & 3  &  14  & 0.5 & 0.833 & 0 & 0 & 0 & 0 & 	 1.7538  &  -5.91  &  0.6  &  9.6  &  15.3    &  14.24  \\  
39 & 3  &  14  & 0.5 & 0.167 & 0 & 0 & 0 & 0 & 	 1.7348  &  -4.24  &  1.1  &  8.8  &  20.6    &  18.69  \\  
40 & 3  &  14  & 0.5 & 0.5 & 0 & 0 & 0 & 0 & 	 1.7427  &  -4.69  &  2.1  &  9.6  &  18.1    &  16.54  \\  
41 & 3  &  14  & 0.5 & 1 & 0 & 0 & 0 & 0 & 	 1.7559  &  -6.15  &  0.3  &  9.9  &  14.9    &  13.88  \\  
42 & 3  &  14  & 0.5 & 0 & 0 & 0 & 0 & 0 & 	 1.7338  &  -3.95  &  0.5  &  8.7  &  21.0    &  18.99  \\  
43 & 3  &  16  & 0.5 & 0.5 & 0 & 0 & 0 & 0 & 	 1.4420  &  -2.83  &  3.4  &  10.  &  26.1    &  24.63  \\  
44 & 3  &  16  & 0.5 & 1 & 0 & 0 & 0 & 0 & 	 1.4494  &  -4.70  &  29.8 &  10.  &  22.1    &  21.13  \\  
45 & 3  &  17  & 0.5 & 0.5 & 0 & 0 & 0 & 0 & 	 1.3227  &  -2.43  &  3.5  &  8.5  &  30.7    &  29.33  \\  
46 & 3  &  17  & 0.5 & 0 & 0 & 0 & 0 & 0 & 	 1.3175  &  -2.00  &  3.1  &  6.1  &  33.8    &  32.64  \\  
47 & 3  &  18  & 0.5 & 1 & 0 & 0 & 0 & 0 & 	 1.2251  &  -2.06  &  5.4  &  9.1  &  31.4    &  24.98  \\  
48 & 3  &  14  & 0.5 & 0.5 & 0 & 0.5 & 0.5 & 1 & 	 1.7423  &  -4.18  &  1.7  &  2.7  &  11.4    &  -  \\  
49 & 3  &  14  & 0.5 & 0.5 & 0 & 0.5 & 0.5 & 0 & 	 1.7409  &  -4.27  &  3.0  &  2.7  &  11.4    &  -  \\  
50 & 3  &  14  & 0.5 & 0.5 & 0 & 0.5 & 1 & 0 & 	 1.7445  &  -5.78  &  4.2  &  2.8  &  11.7    &  -   \\  
51 & 3  &  14  & 0.5 & 0.5 & 0 & 0.5 & 0 & 0 & 	 1.7391  &  -3.23  &  4.4  &  2.6  &  11.1    &  -   \\  
52 & 3  &  14  & 0.5 & 1 & 0 & 0.5 & 1 & 0 & 	 1.7583  &  -7.93  &  2.7  &  5.9  &  13.6    &  13.23  \\  
53 & 3  &  14  & 0.5 & 0 & 0 & 0.5 & 0.5 & 1 & 	 1.7325  &  -3.86  &  1.7  &  5.6  &  19.6    &  18.99  \\  
54 & 3  &  14  & 0.5 & 0 & 0 & 0.5 & 0.5 & 0 & 	 1.7324  &  -3.73  &  1.8  &  5.6  &  19.6    &  18.98  \\  
55 & 3  &  14  & 0.5 & 0 & 0 & 0.5 & 1 & 0 & 	 1.7346  &  -6.27  &  6.7  &  7.7  &  19.9    &  18.34  \\  
56 & 3  &  14  & 0.5 & 0 & 0 & 0.5 & 0 & 0 & 	 1.7302  &  -2.86  &  4.9  &  5.3  &  19.9    &  19.56  \\  
57 & 5  &  12  & 0 & 0 & 0 & 0 & 0 & 0 & 	 2.1686  &  -5.96  &  37.7  &  9.1  &  14.9    &  13.32  \\  
58 & 5  &  14  & 0 & 0 & 0 & 0 & 0 & 0 & 	 1.7433  &  -3.50  &  2.5  &  9.7  &  23.4    &  21.69  \\  
59 & 5  &  15  & 0 & 0 & 0 & 0 & 0 & 0 & 	 1.5809  &  -2.84  &  4.9  &  8.3  &  28.3    &  26.78  \\  
60 & 5  &  9.5 & 0.323 & 0.663 & -0.176 & 0 & 0 & 0 & 	 3.0132  &  -19.9  &  48.2  &  5.9  &  5.6    &  4.99  \\  
61 & 5  &  9.5 & 0.483 & 0.642 & -0.212 & 0 & 0 & 0 & 	 3.0132  &  -19.9  &  0.1  &  8.5  &  4.7    &  4.73  \\  
62 & 5  &  9.5 & 0.5 & 0.644 & -0.213 & 0 & 0 & 0 & 	 3.0132  &  -19.9  &  9.4  &  4.5  &  4.2    &  4.26  \\  
63 & 5  &  12  & 0.5 & 0.644 & -0.213 & 0 & 0 & 0 & 	 2.1755  &  -9.02  &  3.3  &  2.0  &  7.8    &  7.75  \\  
64 & 5  &  15  & 0.5 & 0.5 & 0 & 0 & 0 & 0 & 	 1.5787  &  -2.87  &  21.6  &  2.1  &  13.5    &  -  \\  
65 & 5  &  15  & 0.5 & 1 & 0 & 0 & 0 & 0 & 	 1.5906  &  -4.38  &  34.2  &  9.4  &  22.9    &  21.97  \\  
66 & 5  &  15  & 0.5 & 0 & 0 & 0 & 0 & 0 & 	 1.5695  &  1.62  &  42.0  &  5.6   &  31.7    &  30.93  \\  
67 & 8  &  13  & 0 & 0 & 0 & 0 & 0 & 0 & 	 1.9345  &  -2.98  &  2.9  &  6.8   &  24.9    &  23.56  \\  
68 & 8  &  13  & 0.5 & 1 & 0 & 0 & 0 & 0 & 	 1.9536  &  -4.21  &  5.1  &  7.0   &  18.7    &  17.99  \\  
69 & 8  &  13  & 0.5 & 0 & 0 & 0 & 0 & 0 & 	 1.9221  &  -4.76  &  37.8  &  6.4  &  31.7    &  29.41  \\  
\hline
\end{longtable}

\begin{longtable}[c]{|c|c|c|ccc|ccc|c|c|c|c|c|c|}
\caption{\label{tab:RunsRandom}  
  {\bf Parameters of the runs evolved in ${\cal S}_{5}$. }
  The first column labels the binary configuration, $q=m_A/m_B$ 
  denotes the mass-ratio, $r/m$, $\Omega_i$ and $\dot a$ denote 
  initial separation, orbital frequency and expansion factor, respectively. 
  $\chi_A$ and $\chi_B$ are the dimensionless spin components of the 
  first and the second hole, and $e_{\Omega}$ is the orbital eccentricity.
  The final orbital frequency is given by $\Omega_f$, and $N_f$ 
  denotes the number of orbits $N_f$ between the initial
  frequency $\Omega_i$ and the final frequency $\Omega_f$.
  The last column denote the number of orbits between the initial orbital frequency until the 
  orbital frequency $0.05/m$.
 }\\
\hline
ScN & $q$ & $r/m$ & $\chi_A$ & $\theta_A/\pi$ & $\phi_A/\pi$ & $\chi_B$ & $\theta_B/\pi$ & $\phi_B/\pi$ & $10^2m\Omega_i$ & $10^{5}\dot a$ & $10^{4}e_\Omega$ & $10^2m\Omega_f$ & $N_f$ & $N_{0.05}$  \\ \hline
\endfirsthead
\multicolumn{15}{l}%
	{\tablename\ \thetable\ -- \textit{Continued}} \\
	\hline
ScN & $q$ & $r/m$ & $\chi_A$ & $\theta_A/\pi$ & $\phi_A/\pi$ & $\chi_B$ & $\theta_B/\pi$ & $\phi_B/\pi$ & $10^2m\Omega_i$ & $10^{5}\dot a$ & $10^{4}e_\Omega$ & $10^2m\Omega_f$ & $N_f$ & $N_{0.05}$  \\ \hline
\endhead
\hline \multicolumn{15}{r}{\textit{Continued}}\\ 
\endfoot
\endlastfoot
70 &  1.07  &  15  & 0.205 & 0.965 & 0.091 & 0.385 & 0.229 & -0.185 &  1.5821  &  -5.98  &  0.8  &  9.5 & 17.21    &  15.83   \\  
71 &  1.08  &  15  & 0.106 & 0.400 & 0.590 & 0.212 & 0.340 & -0.191 &  1.5817  &  -5.40  &  0.7  &  13. & 17.73    &  16.00   \\  
72 &  1.08  &  15  & 0.154 & 0.352 & -0.364 & 0.470 & 0.646 & -0.960 &  1.5838  &  -4.56  &  0.2  &  6.5 & 16.09   &  15.37   \\  
73 &  1.10  &  15  & 0.494 & 0.818 & -0.028 & 0.142 & 0.152 & -0.364 &  1.5854  &  -5.85  &  0.5  &  6.0 & 15.40   &  14.88   \\  
74 &  1.12  &  15  & 0.269 & 0.083 & 0.507 & 0.076 & 0.474 & -0.061 &  1.5795  &  -4.35  &  0.5  &  12. & 18.18    &  16.41   \\  
75 &  1.12  &  15  & 0.377 & 0.109 & -0.712 & 0.458 & 0.725 & -0.935 &  1.5809  &  -3.45  &  0.5  &  8.1 & 17.18   &  15.97   \\  
76 &  1.12  &  15  & 0.344 & 0.028 & 0.492 & 0.352 & 0.240 & -0.931 &  1.5761  &  -2.79  &  0.5  &  11. & 19.12    &  17.19   \\  
77 &  1.12  &  15  & 0.243 & 0.517 & -0.496 & 0.269 & 0.625 & -0.792 &  1.5866  &  -4.26  &  2.0  &  11. & 16.91   &  15.43   \\  
78 &  1.17  &  15  & 0.481 & 0.958 & 0.542 & 0.370 & 0.943 & 0.220 &  1.5919  &  -7.31  &  2.3  &  8.6 & 14.44    &  13.57   \\  
79 &  1.26  &  15  & 0.267 & 0.415 & -0.156 & 0.311 & 0.128 & -0.386 &  1.5797  &  -4.97  &  1.0  &  11. & 18.42   &  16.66  \\  
80 &  1.27  &  15  & 0.073 & 0.786 & 0.362 & 0.463 & 0.452 & 0.106 &  1.5829  &  -1.75  &   0.5  &  7.5 & 16.88    &  15.84  \\  
81 &  1.29  &  15  & 0.357 & 0.424 & -0.907 & 0.452 & 0.345 & 0.677 &  1.5839  &  -6.09  &  2.9  &  7.8 & 17.82    &  16.59  \\  
82 &  1.34  &  15  & 0.079 & 0.568 & -0.523 & 0.200 & 0.853 & -0.820 &  1.5844  &  -5.45  &  0.3  &  11. & 16.95   &  15.53  \\  
83 &  1.35  &  15  & 0.140 & 0.094 & -0.122 & 0.413 & 0.091 & 0.430 &  1.5777  &  -3.31  &  0.2  &  10. & 18.91    &  17.18  \\  
84 &  1.36  &  15  & 0.089 & 0.005 & -0.975 & 0.356 & 0.568 & -0.404 &  1.5866  &  -6.51  &  2.9  &  11. & 17.69   &  16.09  \\  
85 &  1.38  &  15  & 0.169 & 0.531 & -0.097 & 0.455 & 0.235 & -0.152 &  1.5799  &  -5.88  &  1.0  &  7.9 & 17.89   &  16.67  \\  
86 &  1.55  &  15  & 0.082 & 0.820 & -0.171 & 0.205 & 0.825 & -0.052 &  1.5847  &  -5.74  &  0.4  &  11. & 17.20   &  15.76  \\  
87 &  1.63  &  15  & 0.114 & 0.521 & -0.665 & 0.325 & 0.101 & -0.587 &  1.5806  &  -3.49  &  1.2  &  9.8 & 18.57  &  17.05   \\  
88 &  1.63  &  15  & 0.295 & 0.516 & 0.635 & 0.179 & 0.060 & 0.310 &  1.5832  &  -5.15  &  1.7  &  12. & 18.50    &  16.78   \\  
89 &  1.63  &  15  & 0.171 & 0.786 & -0.032 & 0.155 & 0.544 & -0.983 &  1.5843  &  -5.15  &  0.2  &  11. & 17.46   &  15.99  \\  
90 &  1.64  &  15  & 0.115 & 0.435 & 0.808 & 0.343 & 0.730 & -0.361 &  1.5861  &  -6.82  &  1.4  &  11. &  17.65  &  16.17   \\  
91 &  1.68  &  15  & 0.365 & 0.148 & -0.283 & 0.217 & 0.269 & -0.911 &  1.5772  &  -3.53  &  0.8  &  12. &  20.04  &  18.03  \\  
92 &  1.70  &  15  & 0.429 & 0.470 & 0.487 & 0.112 & 0.958 & -0.874 &  1.5880  &  -5.20  &  4.9  &  12. &  18.22  &  16.60   \\  
93 & 1.76  &  15  & 0.406 & 0.530 & -0.810 & 0.252 & 0.764 & 0.476 &  1.5866  &  -2.88  &  1.5  &  11. &  17.79   &  16.33   \\  
94 & 1.82  &  15  & 0.403 & 0.326 & -0.511 & 0.146 & 0.590 & -0.953 &  1.5828  &  -4.13  &    2.4  &  11. &  19.44 &  17.67  \\  
95 & 1.82  &  15  & 0.438 & 0.658 & -0.446 & 0.124 & 0.253 & 0.690 &  1.5891  &  -5.74  &  2.9  &  9.5 & 17.56    &  16.29   \\  
96 & 1.84  &  15  & 0.208 & 0.364 & 0.875 & 0.160 & 0.420 & 0.640 &  1.5815  &  -4.99  &  0.9  &  11. & 19.08    &   17.40   \\  
97 & 1.91  &  15  & 0.375 & 0.545 & -0.854 & 0.110 & 0.012 & -0.107 &  1.5826  &  -2.71  &    0.6  &  11. & 18.76  &  17.17  \\  
98 & 1.92  &  15  & 0.451 & 0.082 & -0.887 & 0.443 & 0.563 & -0.652 &  1.5834  &  +3.02  &  4.6  &  8.6 & 20.34    &  18.73  \\  
99 & 1.94  &  15  & 0.312 & 0.873 & 0.292 & 0.222 & 0.080 & -0.934 &  1.5853  &  -4.26  &  0.8  &  11. & 17.88     &  16.46  \\  
100 & 1.96  &  15  & 0.057 & 0.810 & 0.274 & 0.090 & 0.408 & -0.113 &  1.5827  &  -4.68  &  0.4  &  12. & 18.76    &  17.14  \\  
101 & 1.97  &  15  & 0.190 & 0.564 & -0.917 & 0.372 & 0.911 & 0.727 &  1.5851  &  -5.78  &  0.5  &  9.5 &  17.82   &  16.56 \\ \hline
\end{longtable}


\begingroup
\begin{table}[h]
\parbox{0.45\textwidth}{
\caption{\label{tab:aij}  
  {\bf Parameters $a_{i,j}$ of $\rho_{\rm NS}$.}
}
\begin{tabular}{|c|c|c|c|c|}
\hline
$i$ & $j=0$ & $j=1$ & $j=2$ & $j=3$
\\\hline
0 & -2.23773 & 89.0053 & -907.744 & 3194.70 \\
1 & 16.3786 & -321.623 & 1589.74 & 0 \\
2 & -25.8886 & 285.810 & 0 & 0 \\
3 & 7.94034 & 0 & 0 & 0 \\
\hline 
\end{tabular}
}
%
\parbox{0.45\textwidth}{
\caption{\label{tab:bij}  
  {\bf Parameters $b_{i,j}$ of $\kappa_{\rm NS}$.}
}
\begin{tabular}{|c|c|c|c|c|}
\hline
$i$ & $j=0$ & $j=1$ & $j=2$ & $j=3$
\\\hline
0 & 0.999768 & -2.63274 & 4.08714 & -2.67330 \\
1 & -0.0126939 & 0.343292 & 1.75169 & 0 \\
2 & 0.0445636 & -0.975111 & 0 & 0 \\
3 & 0.0157333 & 0 & 0 & 0 \\
\hline 
\end{tabular}
}
\end{table}
\endgroup

\begin{longtable}{|c|c|c|c|c|}
\caption{\label{tab:cijk}  
  {\bf Parameters $c_{i,j,k}$ of $\delta\kappa_{\rm S}$.}
}
\\
\hline
$(i,k)$ & $j=1$ & $j=2$ & $j=3$ & $j=4$ 
\\\hline
\endfirsthead
\multicolumn{5}{l}
{\tablename\ \thetable\ -- \textit{Continued}} \\
\hline
$(i,k)$ & $j=1$ & $j=2$ & $j=3$ & $j=4$ 
\\\hline
\endhead
\hline \multicolumn{5}{r}{\textit{Continued}}\\
\endfoot
\endlastfoot
(0,1) & 5316.71 & -66338.2 & 275375. & -380348. \\
(0,2) & 32067.7 & -376132. & 1.48805$\times 10^6$ & -1.93024$\times 10^6$ \\
(0,3) & 14274.6 & -177525. & 736095. & -1.01733$\times 10^6$ \\
(0,4) & -1843.34 & 9864.44 & 1329.56 & -4958.02 \\
(0,5) & 109.978 & 517.973 & -7645.17 & 15290.1 \\
(0,6) & 23169.4 & -275526. & 1.08011$\times 10^6$ & -1.42663$\times 10^6$ \\
(0,7) & 7819.45 & -101310. & 434622. & -617796. \\
(0,8) & -42885.3 & 532586. & -2.20338$\times 10^6$ & 3.03600$\times 10^6$ \\
(0,9) & 1027.50 & -27767.7 & 103643. & -218964. \\
(0,10) & -109193. & 1.29526$\times 10^6$ & -5.05535$\times 10^6$ & 6.62736$\times 10^6$ \\
(0,11) & -55234.7 & 704761. & -2.98370$\times 10^6$ & 4.19269$\times 10^6$ \\
(0,12) & -92749.8 & 1.02247$\times 10^6$ & -4.02065$\times 10^6$ & 5.29669$\times 10^6$ \\
(0,13) & -7921.47 & 97674.7 & -405463. & 565779. \\
(0,14) & -103301. & 1.22607$\times 10^6$ & -4.82157$\times 10^6$ & 6.34451$\times 10^6$ \\
(1,1) & -82960.2 & 1.03434$\times 10^6$ & -4.29063$\times 10^6$ & 5.92217$\times 10^6$ \\
(1,2) & -496416. & 5.85007$\times 10^6$ & -2.32014$\times 10^7$ & 3.02622$\times 10^7$ \\
(1,3) & -217655. & 2.71089$\times 10^6$ & -1.12563$\times 10^7$ & 1.55773$\times 10^7$ \\
(1,4) & 25018.3 & -393524. & 2.05655$\times 10^6$ & -4.12184$\times 10^6$ \\
(1,5) & 28030.5 & -368855. & 1.57451$\times 10^6$ & -2.19102$\times 10^6$ \\
(1,6) & -412124. & 4.87348$\times 10^6$ & -1.90352$\times 10^7$ & 2.49538$\times 10^7$ \\
(1,7) & -128584. & 1.65610$\times 10^6$ & -7.06852$\times 10^6$ & 1.00034$\times 10^7$ \\
(1,8) & 661241. & -8.21574$\times 10^6$ & 3.40040$\times 10^7$ & -4.68715$\times 10^7$ \\
(1,9) & -24548.5 & 478963. & -1.79194$\times 10^6$ & 3.37544$\times 10^6$ \\
(1,10) & 1.69467$\times 10^6$ & -2.01047$\times 10^7$ & 7.86851$\times 10^7$ & -1.03283$\times 10^8$ \\
(1,11) & 917140. & -1.16445$\times 10^7$ & 4.90819$\times 10^7$ & -6.86985$\times 10^7$ \\
(1,12) & -289050. & 3.65366$\times 10^6$ & -1.08819$\times 10^7$ & 8.95028$\times 10^6$ \\
(1,13) & 46482.3 & -583937. & 2.51586$\times 10^6$ & -3.69108$\times 10^6$ \\
(1,14) & 1.74966$\times 10^6$ & -2.07098$\times 10^7$ & 8.12790$\times 10^7$ & -1.06573$\times 10^8$ \\
(2,1) & 424444. & -5.28782$\times 10^6$ & 2.19181$\times 10^7$ & -3.02304$\times 10^7$ \\
(2,2) & 2.54951$\times 10^6$ & -3.01226$\times 10^7$ & 1.19551$\times 10^8$ & -1.56397$\times 10^8$ \\
(2,3) & 1.08832$\times 10^6$ & -1.35750$\times 10^7$ & 5.64446$\times 10^7$ & -7.82131$\times 10^7$ \\
(2,4) & 538865. & -4.65423$\times 10^6$ & 1.08665$\times 10^7$ & -650921. \\
(2,5) & -134766. & 1.75054$\times 10^6$ & -7.36970$\times 10^6$ & 1.00998$\times 10^7$ \\
(2,6) & 2.38473$\times 10^6$ & -2.81363$\times 10^7$ & 1.09819$\times 10^8$ & -1.43508$\times 10^8$ \\
(2,7) & 690728. & -8.84984$\times 10^6$ & 3.76020$\times 10^7$ & -5.30045$\times 10^7$ \\
(2,8) & -3.34460$\times 10^6$ & 4.15737$\times 10^7$ & -1.72135$\times 10^8$ & 2.37354$\times 10^8$ \\
(2,9) & 149314. & -2.61396$\times 10^6$ & 1.01824$\times 10^7$ & -1.80892$\times 10^7$ \\
(2,10) & -8.70071$\times 10^6$ & 1.03298$\times 10^8$ & -4.05413$\times 10^8$ & 5.32989$\times 10^8$ \\
(2,11) & -4.98684$\times 10^6$ & 6.30344$\times 10^7$ & -2.64631$\times 10^8$ & 3.69064$\times 10^8$ \\
(2,12) & 362328. & 2.89435$\times 10^6$ & -6.26526$\times 10^7$ & 1.54943$\times 10^8$ \\
(2,13) & -296859. & 3.76776$\times 10^6$ & -1.62996$\times 10^7$ & 2.38893$\times 10^7$ \\
(2,14) & -9.73227$\times 10^6$ & 1.15044$\times 10^8$ & -4.51208$\times 10^8$ & 5.90608$\times 10^8$ \\
(3,1) & -709184. & 8.82898$\times 10^6$ & -3.65705$\times 10^7$ & 5.04053$\times 10^7$ \\
(3,2) & -4.35305$\times 10^6$ & 5.15001$\times 10^7$ & -2.04343$\times 10^8$ & 2.67729$\times 10^8$ \\
(3,3) & -1.78215$\times 10^6$ & 2.22596$\times 10^7$ & -9.26725$\times 10^7$ & 1.28564$\times 10^8$ \\
(3,4) & 908964. & -1.60025$\times 10^7$ & 8.51391$\times 10^7$ & -1.47147$\times 10^8$ \\
(3,5) & 62120.2 & -864159. & 3.64740$\times 10^6$ & -4.74819$\times 10^6$ \\
(3,6) & -4.48307$\times 10^6$ & 5.28627$\times 10^7$ & -2.06464$\times 10^8$ & 2.69548$\times 10^8$ \\
(3,7) & -1.20683$\times 10^6$ & 1.53949$\times 10^7$ & -6.51614$\times 10^7$ & 9.15436$\times 10^7$ \\
(3,8) & 5.53612$\times 10^6$ & -6.88414$\times 10^7$ & 2.85135$\times 10^8$ & -3.93290$\times 10^8$ \\
(3,9) & -265327. & 4.52185$\times 10^6$ & -1.85463$\times 10^7$ & 3.22321$\times 10^7$ \\
(3,10) & 1.48479$\times 10^7$ & -1.76472$\times 10^8$ & 6.94463$\times 10^8$ & -9.14535$\times 10^8$ \\
(3,11) & 8.83120$\times 10^6$ & -1.11213$\times 10^8$ & 4.65325$\times 10^8$ & -6.46984$\times 10^8$ \\
(3,12) & -6.70535$\times 10^6$ & 5.32904$\times 10^7$ & -6.84880$\times 10^7$ & -1.07762$\times 10^8$ \\
(3,13) & 1.00833$\times 10^6$ & -1.27036$\times 10^7$ & 5.39248$\times 10^7$ & -7.69297$\times 10^7$ \\
(3,14) & 1.77463$\times 10^7$ & -2.09671$\times 10^8$ & 8.22363$\times 10^8$ & -1.07571$\times 10^9$ \\
\hline 
\end{longtable}

\begin{longtable}{|c|c|c|c|c|}
\caption{\label{tab:dijk}  
  {\bf Parameters $d_{i,j,k}$ of $\delta\rho_{\rm S}$.}
}
\\
\hline
$(i,k)$ & $j=1$ & $j=2$ & $j=3$ & $j=4$ 
\\\hline
\endfirsthead
\multicolumn{5}{l}
{\tablename\ \thetable\ -- \textit{Continued}} \\
\hline
$(i,k)$ & $j=1$ & $j=2$ & $j=3$ & $j=4$ 
\\\hline
\endhead
\hline \multicolumn{5}{r}{\textit{Continued}}\\
\endfoot
\endlastfoot
(0,1) & -45382.3 & 302204. & -153169. & -1.30349$\times 10^6$ \\
(0,2) & -253943. & 1.90739$\times 10^6$ & 632048. & -7.50294$\times 10^6$ \\
(0,3) & -1.31299$\times 10^6$ & 1.54603$\times 10^7$ & -6.03927$\times 10^7$ & 7.82728$\times 10^7$ \\
(0,4) & -1.20248$\times 10^6$ & 8.98755$\times 10^6$ & -2.09367$\times 10^7$ & 2.51332$\times 10^7$ \\
(0,5) & 550699. & -6.20436$\times 10^6$ & 2.42440$\times 10^7$ & -3.28133$\times 10^7$ \\
(0,6) & 587207. & -7.79345$\times 10^6$ & 3.28891$\times 10^7$ & -4.78587$\times 10^7$ \\
(0,7) & -234908. & 2.08401$\times 10^6$ & -5.31451$\times 10^6$ & 2.96473$\times 10^6$ \\
(0,8) & 529948. & -4.20533$\times 10^6$ & 7.57153$\times 10^6$ & 2.97058$\times 10^6$ \\
(0,9) & 677458. & -1.23094$\times 10^7$ & 5.30880$\times 10^7$ & -9.37766$\times 10^7$ \\
(0,10) & 4.56142$\times 10^6$ & -5.15215$\times 10^7$ & 2.05792$\times 10^8$ & -2.57620$\times 10^8$ \\
(0,11) & -5.37955$\times 10^6$ & 8.86716$\times 10^7$ & -4.55622$\times 10^8$ & 7.47340$\times 10^8$ \\
(0,12) & -1.23435$\times 10^7$ & 1.01134$\times 10^8$ & -3.15628$\times 10^8$ & 3.51274$\times 10^8$ \\
(0,13) & -1.45409$\times 10^6$ & 1.69206$\times 10^7$ & -6.86296$\times 10^7$ & 9.64437$\times 10^7$ \\
(0,14) & 5.15330$\times 10^6$ & -5.79718$\times 10^7$ & 2.18460$\times 10^8$ & -2.69339$\times 10^8$ \\
(1,1) & -45382.3 & 302204. & -153169. & -1.30349$\times 10^6$ \\
(1,2) & -253943. & 1.90739$\times 10^6$ & 632048. & -7.50294$\times 10^6$ \\
(1,3) & -1.31299$\times 10^6$ & 1.54603$\times 10^7$ & -6.03927$\times 10^7$ & 7.82728$\times 10^7$ \\
(1,4) & -1.20248$\times 10^6$ & 8.98755$\times 10^6$ & -2.09367$\times 10^7$ & 2.51332$\times 10^7$ \\
(1,5) & 550699. & -6.20436$\times 10^6$ & 2.42440$\times 10^7$ & -3.28133$\times 10^7$ \\
(1,6) & 587207. & -7.79345$\times 10^6$ & 3.28891$\times 10^7$ & -4.78587$\times 10^7$ \\
(1,7) & -234908. & 2.08401$\times 10^6$ & -5.31451$\times 10^6$ & 2.96473$\times 10^6$ \\
(1,8) & 529948. & -4.20533$\times 10^6$ & 7.57153$\times 10^6$ & 2.97058$\times 10^6$ \\
(1,9) & 677458. & -1.23094$\times 10^7$ & 5.30880$\times 10^7$ & -9.37766$\times 10^7$ \\
(1,10) & 4.56142$\times 10^6$ & -5.15215$\times 10^7$ & 2.05792$\times 10^8$ & -2.57620$\times 10^8$ \\
(1,11) & -5.37955$\times 10^6$ & 8.86716$\times 10^7$ & -4.55622$\times 10^8$ & 7.47340$\times 10^8$ \\
(1,12) & -1.23435$\times 10^7$ & 1.01134$\times 10^8$ & -3.15628$\times 10^8$ & 3.51274$\times 10^8$ \\
(1,13) & -1.45409$\times 10^6$ & 1.69206$\times 10^7$ & -6.86296$\times 10^7$ & 9.64437$\times 10^7$ \\
(1,14) & 5.15330$\times 10^6$ & -5.79718$\times 10^7$ & 2.18460$\times 10^8$ & -2.69339$\times 10^8$ \\
(2,1) & -45382.3 & 302204. & -153169. & -1.30349$\times 10^6$ \\
(2,2) &  -253943. & 1.90739$\times 10^6$ & 632048. & -7.50294$\times 10^6$ \\
(2,3) &  -1.31299$\times 10^6$ & 1.54603$\times 10^7$ & -6.03927$\times 10^7$ & 7.82728$\times 10^7$ \\
(2,4) &  -1.20248$\times 10^6$ & 8.98755$\times 10^6$ & -2.09367$\times 10^7$ & 2.51332$\times 10^7$ \\
(2,5) &  550699. & -6.20436$\times 10^6$ & 2.42440$\times 10^7$ & -3.28133$\times 10^7$ \\
(2,6) &  587207. & -7.79345$\times 10^6$ & 3.28891$\times 10^7$ & -4.78587$\times 10^7$ \\
(2,7) & -234908. & 2.08401$\times 10^6$ & -5.31451$\times 10^6$ & 2.96473$\times 10^6$ \\
(2,8) &  529948. & -4.20533$\times 10^6$ & 7.57153$\times 10^6$ & 2.97058$\times 10^6$ \\
(2,9) &  677458. & -1.23094$\times 10^7$ & 5.30880$\times 10^7$ & -9.37766$\times 10^7$ \\
(2,10) &  4.56142$\times 10^6$ & -5.15215$\times 10^7$ & 2.05792$\times 10^8$ & -2.57620$\times 10^8$ \\
(2,11) &  -5.37955$\times 10^6$ & 8.86716$\times 10^7$ & -4.55622$\times 10^8$ & 7.47340$\times 10^8$ \\
(2,12) &  -1.23435$\times 10^7$ & 1.01134$\times 10^8$ & -3.15628$\times 10^8$ & 3.51274$\times 10^8$ \\
(2,13) &  -1.45409$\times 10^6$ & 1.69206$\times 10^7$ & -6.86296$\times 10^7$ & 9.64437$\times 10^7$ \\
(2,14) &  5.15330$\times 10^6$ & -5.79718$\times 10^7$ & 2.18460$\times 10^8$ & -2.69339$\times 10^8$ \\
(3,1) & -45382.3 & 302204. & -153169. & -1.30349$\times 10^6$ \\
(3,2) & -253943. & 1.90739$\times 10^6$ & 632048. & -7.50294$\times 10^6$ \\
(3,3) & -1.31299$\times 10^6$ & 1.54603$\times 10^7$ & -6.03927$\times 10^7$ & 7.82728$\times 10^7$ \\
(3,4) & -1.20248$\times 10^6$ & 8.98755$\times 10^6$ & -2.09367$\times 10^7$ & 2.51332$\times 10^7$ \\
(3,5) & 550699. & -6.20436$\times 10^6$ & 2.42440$\times 10^7$ & -3.28133$\times 10^7$ \\
(3,6) & 587207. & -7.79345$\times 10^6$ & 3.28891$\times 10^7$ & -4.78587$\times 10^7$ \\
(3,7) & -234908. & 2.08401$\times 10^6$ & -5.31451$\times 10^6$ & 2.96473$\times 10^6$ \\
(3,8) & 529948. & -4.20533$\times 10^6$ & 7.57153$\times 10^6$ & 2.97058$\times 10^6$ \\
(3,9) & 677458. & -1.23094$\times 10^7$ & 5.30880$\times 10^7$ & -9.37766$\times 10^7$ \\
(3,10) & 4.56142$\times 10^6$ & -5.15215$\times 10^7$ & 2.05792$\times 10^8$ & -2.57620$\times 10^8$ \\
(3,11) & -5.37955$\times 10^6$ & 8.86716$\times 10^7$ & -4.55622$\times 10^8$ & 7.47340$\times 10^8$ \\
(3,12) & -1.23435$\times 10^7$ & 1.01134$\times 10^8$ & -3.15628$\times 10^8$ & 3.51274$\times 10^8$ \\
(3,13) & -1.45409$\times 10^6$ & 1.69206$\times 10^7$ & -6.86296$\times 10^7$ & 9.64437$\times 10^7$ \\
(3,14) & 5.15330$\times 10^6$ & -5.79718$\times 10^7$ & 2.18460$\times 10^8$ & -2.69339$\times 10^8$ \\
\hline 
\end{longtable}

\end{widetext}

\bibliography{References}

\begin{thebibliography}{101}
\expandafter\ifx\csname natexlab\endcsname\relax\def\natexlab#1{#1}\fi
\expandafter\ifx\csname bibnamefont\endcsname\relax
  \def\bibnamefont#1{#1}\fi
\expandafter\ifx\csname bibfnamefont\endcsname\relax
  \def\bibfnamefont#1{#1}\fi
\expandafter\ifx\csname citenamefont\endcsname\relax
  \def\citenamefont#1{#1}\fi
\expandafter\ifx\csname url\endcsname\relax
  \def\url#1{\texttt{#1}}\fi
\expandafter\ifx\csname urlprefix\endcsname\relax\def\urlprefix{URL }\fi
\providecommand{\bibinfo}[2]{#2}
\providecommand{\eprint}[2][]{\url{#2}}

\bibitem[{Adv()}]{AdvancedLIGOWebsite}
\emph{\bibinfo{title}{Advanced ligo}},
  \bibinfo{note}{http://lhocds.ligo-wa.caltech.edu:8000/advligo/GWINC}.

\bibitem[{\citenamefont{scientific collaboration}()}]{LIGOScience}
\bibinfo{author}{\bibfnamefont{L.}~\bibnamefont{scientific collaboration}},
  \emph{\bibinfo{title}{The science of lsc research}},
  \urlprefix\url{http://www.ligo.org/science/overview.php}.

\bibitem[{\citenamefont{Abadie et~al.}(2012)}]{Abadie:2012}
\bibinfo{author}{\bibfnamefont{J.}~\bibnamefont{Abadie}} \bibnamefont{et~al.}
  (\bibinfo{collaboration}{The LIGO Scientific Collaboration and the Virgo
  Collaboration, the Virgo Collaboration}) (\bibinfo{year}{2012}),
  \eprint{1201.5999}.

\bibitem[{\citenamefont{Acernese et~al.}(2008)\citenamefont{Acernese,
  Alshourbagy, Amico, Antonucci, Aoudia et~al.}}]{Acernese:2008}
\bibinfo{author}{\bibfnamefont{F.}~\bibnamefont{Acernese}},
  \bibinfo{author}{\bibfnamefont{M.}~\bibnamefont{Alshourbagy}},
  \bibinfo{author}{\bibfnamefont{P.}~\bibnamefont{Amico}},
  \bibinfo{author}{\bibfnamefont{F.}~\bibnamefont{Antonucci}},
  \bibinfo{author}{\bibfnamefont{S.}~\bibnamefont{Aoudia}},
  \bibnamefont{et~al.}, \bibinfo{journal}{Class.Quant.Grav.}
  \textbf{\bibinfo{volume}{25}}, \bibinfo{pages}{184001}
  (\bibinfo{year}{2008}).

\bibitem[{\citenamefont{Somiya and the
  {KAGRA}~Collaboration}(2012)}]{Somiya:2012}
\bibinfo{author}{\bibfnamefont{K.}~\bibnamefont{Somiya}} \bibnamefont{and}
  \bibinfo{author}{\bibnamefont{the {KAGRA}~Collaboration}},
  \bibinfo{journal}{Class.\ Quantum Grav.} \textbf{\bibinfo{volume}{29}},
  \bibinfo{pages}{124007} (\bibinfo{year}{2012}).

\bibitem[{\citenamefont{Pretorius}(2005)}]{Pretorius2005c}
\bibinfo{author}{\bibfnamefont{F.}~\bibnamefont{Pretorius}},
  \bibinfo{journal}{Class.\ Quantum Grav.} \textbf{\bibinfo{volume}{22}},
  \bibinfo{pages}{425} (\bibinfo{year}{2005}).

\bibitem[{\citenamefont{Centrella et~al.}(2010)\citenamefont{Centrella, Baker,
  Kelly, and van Meter}}]{Centrella:2010}
\bibinfo{author}{\bibfnamefont{J.}~\bibnamefont{Centrella}},
  \bibinfo{author}{\bibfnamefont{J.~G.} \bibnamefont{Baker}},
  \bibinfo{author}{\bibfnamefont{B.~J.} \bibnamefont{Kelly}}, \bibnamefont{and}
  \bibinfo{author}{\bibfnamefont{J.~R.} \bibnamefont{van Meter}},
  \bibinfo{journal}{Rev.\ Mod.\ Phys.} \textbf{\bibinfo{volume}{82}},
  \bibinfo{pages}{3069} (\bibinfo{year}{2010}).

\bibitem[{\citenamefont{Pfeiffer}(2012)}]{Pfeiffer:2012pc}
\bibinfo{author}{\bibfnamefont{H.~P.} \bibnamefont{Pfeiffer}},
  \emph{\bibinfo{title}{{Numerical simulations of compact object binaries}}}
  (\bibinfo{year}{2012}), \eprint{1203.5166}.

\bibitem[{\citenamefont{Buonanno
  et~al.}(2007{\natexlab{a}})\citenamefont{Buonanno, Pan, Baker, Centrella,
  Kelly et~al.}}]{Buonanno2007}
\bibinfo{author}{\bibfnamefont{A.}~\bibnamefont{Buonanno}},
  \bibinfo{author}{\bibfnamefont{Y.}~\bibnamefont{Pan}},
  \bibinfo{author}{\bibfnamefont{J.~G.} \bibnamefont{Baker}},
  \bibinfo{author}{\bibfnamefont{J.}~\bibnamefont{Centrella}},
  \bibinfo{author}{\bibfnamefont{B.~J.} \bibnamefont{Kelly}},
  \bibnamefont{et~al.}, \bibinfo{journal}{Phys.\ Rev.\ D}
  \textbf{\bibinfo{volume}{76}}, \bibinfo{pages}{104049}
  (\bibinfo{year}{2007}{\natexlab{a}}), \eprint{0706.3732}.

\bibitem[{\citenamefont{Ajith et~al.}(2008)\citenamefont{Ajith, Babak, Chen,
  Hewitson, Krishnan et~al.}}]{Ajith-Babak-Chen-etal:2007b}
\bibinfo{author}{\bibfnamefont{P.}~\bibnamefont{Ajith}},
  \bibinfo{author}{\bibfnamefont{S.}~\bibnamefont{Babak}},
  \bibinfo{author}{\bibfnamefont{Y.}~\bibnamefont{Chen}},
  \bibinfo{author}{\bibfnamefont{M.}~\bibnamefont{Hewitson}},
  \bibinfo{author}{\bibfnamefont{B.}~\bibnamefont{Krishnan}},
  \bibnamefont{et~al.}, \bibinfo{journal}{Phys.\ Rev.\ D}
  \textbf{\bibinfo{volume}{77}}, \bibinfo{pages}{104017}
  (\bibinfo{year}{2008}), \eprint{0710.2335}.

\bibitem[{\citenamefont{Damour and Nagar}(2009)}]{Damour2009a}
\bibinfo{author}{\bibfnamefont{T.}~\bibnamefont{Damour}} \bibnamefont{and}
  \bibinfo{author}{\bibfnamefont{A.}~\bibnamefont{Nagar}},
  \bibinfo{journal}{Phys.\ Rev.\ D} \textbf{\bibinfo{volume}{79}},
  \bibinfo{pages}{081503} (\bibinfo{year}{2009}), \eprint{0902.0136}.

\bibitem[{\citenamefont{Buonanno et~al.}(2009)\citenamefont{Buonanno, Pan,
  Pfeiffer, Scheel, Buchman et~al.}}]{Buonanno:2009qa}
\bibinfo{author}{\bibfnamefont{A.}~\bibnamefont{Buonanno}},
  \bibinfo{author}{\bibfnamefont{Y.}~\bibnamefont{Pan}},
  \bibinfo{author}{\bibfnamefont{H.~P.} \bibnamefont{Pfeiffer}},
  \bibinfo{author}{\bibfnamefont{M.~A.} \bibnamefont{Scheel}},
  \bibinfo{author}{\bibfnamefont{L.~T.} \bibnamefont{Buchman}},
  \bibnamefont{et~al.}, \bibinfo{journal}{Phys.\ Rev.\ D}
  \textbf{\bibinfo{volume}{79}}, \bibinfo{pages}{124028}
  (\bibinfo{year}{2009}), \eprint{0902.0790}.

\bibitem[{\citenamefont{Ajith et~al.}(2009{\natexlab{a}})\citenamefont{Ajith,
  Hannam, Husa, Chen, Bruegmann, Dorband, Mueller, Ohme, Pollney, Reisswig
  et~al.}}]{Ajith2009}
\bibinfo{author}{\bibfnamefont{P.}~\bibnamefont{Ajith}},
  \bibinfo{author}{\bibfnamefont{M.}~\bibnamefont{Hannam}},
  \bibinfo{author}{\bibfnamefont{S.}~\bibnamefont{Husa}},
  \bibinfo{author}{\bibfnamefont{Y.}~\bibnamefont{Chen}},
  \bibinfo{author}{\bibfnamefont{B.}~\bibnamefont{Bruegmann}},
  \bibinfo{author}{\bibfnamefont{N.}~\bibnamefont{Dorband}},
  \bibinfo{author}{\bibfnamefont{D.}~\bibnamefont{Mueller}},
  \bibinfo{author}{\bibfnamefont{F.}~\bibnamefont{Ohme}},
  \bibinfo{author}{\bibfnamefont{D.}~\bibnamefont{Pollney}},
  \bibinfo{author}{\bibfnamefont{C.}~\bibnamefont{Reisswig}},
  \bibnamefont{et~al.} (\bibinfo{year}{2009}{\natexlab{a}}),
  \eprint{0909.2867}.

\bibitem[{\citenamefont{Pan et~al.}(2010)\citenamefont{Pan, Buonanno, Buchman,
  Chu, Kidder et~al.}}]{Pan:2009wj}
\bibinfo{author}{\bibfnamefont{Y.}~\bibnamefont{Pan}},
  \bibinfo{author}{\bibfnamefont{A.}~\bibnamefont{Buonanno}},
  \bibinfo{author}{\bibfnamefont{L.~T.} \bibnamefont{Buchman}},
  \bibinfo{author}{\bibfnamefont{T.}~\bibnamefont{Chu}},
  \bibinfo{author}{\bibfnamefont{L.~E.} \bibnamefont{Kidder}},
  \bibnamefont{et~al.}, \bibinfo{journal}{Phys.Rev.}
  \textbf{\bibinfo{volume}{D81}}, \bibinfo{pages}{084041}
  (\bibinfo{year}{2010}), \eprint{0912.3466}.

\bibitem[{\citenamefont{Aylott et~al.}(2009{\natexlab{a}})}]{ninjashort}
\bibinfo{author}{\bibfnamefont{B.}~\bibnamefont{Aylott}} \bibnamefont{et~al.},
  \bibinfo{journal}{Class.\ Quantum Grav.} \textbf{\bibinfo{volume}{26}},
  \bibinfo{pages}{165008} (\bibinfo{year}{2009}{\natexlab{a}}).

\bibitem[{\citenamefont{Santamar\'{i}a
  et~al.}(2010)\citenamefont{Santamar\'{i}a, Ohme, Ajith, Br{\"u}gmann,
  Dorband, Hannam, Husa, M{\"o}sta, Pollney, Reisswig
  et~al.}}]{Santamaria:2010yb}
\bibinfo{author}{\bibfnamefont{L.}~\bibnamefont{Santamar\'{i}a}},
  \bibinfo{author}{\bibfnamefont{F.}~\bibnamefont{Ohme}},
  \bibinfo{author}{\bibfnamefont{P.}~\bibnamefont{Ajith}},
  \bibinfo{author}{\bibfnamefont{B.}~\bibnamefont{Br{\"u}gmann}},
  \bibinfo{author}{\bibfnamefont{N.}~\bibnamefont{Dorband}},
  \bibinfo{author}{\bibfnamefont{M.}~\bibnamefont{Hannam}},
  \bibinfo{author}{\bibfnamefont{S.}~\bibnamefont{Husa}},
  \bibinfo{author}{\bibfnamefont{P.}~\bibnamefont{M{\"o}sta}},
  \bibinfo{author}{\bibfnamefont{D.}~\bibnamefont{Pollney}},
  \bibinfo{author}{\bibfnamefont{C.}~\bibnamefont{Reisswig}},
  \bibnamefont{et~al.}, \bibinfo{journal}{Phys.\ Rev.\ D}
  \textbf{\bibinfo{volume}{82}}, \bibinfo{pages}{064016}
  (\bibinfo{year}{2010}).

\bibitem[{\citenamefont{Ajith et~al.}(2012)\citenamefont{Ajith, Boyle, Brown,
  Brugmann, Buchman et~al.}}]{Ajith:2012tt}
\bibinfo{author}{\bibfnamefont{P.}~\bibnamefont{Ajith}},
  \bibinfo{author}{\bibfnamefont{M.}~\bibnamefont{Boyle}},
  \bibinfo{author}{\bibfnamefont{D.~A.} \bibnamefont{Brown}},
  \bibinfo{author}{\bibfnamefont{B.}~\bibnamefont{Brugmann}},
  \bibinfo{author}{\bibfnamefont{L.~T.} \bibnamefont{Buchman}},
  \bibnamefont{et~al.}, \bibinfo{journal}{Class.\ Quantum Grav.}
  \textbf{\bibinfo{volume}{29}}, \bibinfo{pages}{124001}
  (\bibinfo{year}{2012}),
  \urlprefix\url{http://stacks.iop.org/0264-9381/29/i=12/a=124001}.

\bibitem[{\citenamefont{Aylott et~al.}(2009{\natexlab{b}})\citenamefont{Aylott,
  Baker, Boggs, Boyle, Brady et~al.}}]{Aylott:2009ya}
\bibinfo{author}{\bibfnamefont{B.}~\bibnamefont{Aylott}},
  \bibinfo{author}{\bibfnamefont{J.~G.} \bibnamefont{Baker}},
  \bibinfo{author}{\bibfnamefont{W.~D.} \bibnamefont{Boggs}},
  \bibinfo{author}{\bibfnamefont{M.}~\bibnamefont{Boyle}},
  \bibinfo{author}{\bibfnamefont{P.~R.} \bibnamefont{Brady}},
  \bibnamefont{et~al.}, \bibinfo{journal}{Class.Quant.Grav.}
  \textbf{\bibinfo{volume}{26}}, \bibinfo{pages}{165008}
  (\bibinfo{year}{2009}{\natexlab{b}}), \eprint{0901.4399}.

\bibitem[{\citenamefont{Peters and Mathews}(1963)}]{PetersMathews1963}
\bibinfo{author}{\bibfnamefont{P.~C.} \bibnamefont{Peters}} \bibnamefont{and}
  \bibinfo{author}{\bibfnamefont{J.}~\bibnamefont{Mathews}},
  \bibinfo{journal}{Phys. Rev.} \textbf{\bibinfo{volume}{131}},
  \bibinfo{pages}{435} (\bibinfo{year}{1963}).

\bibitem[{\citenamefont{Peters}(1964)}]{Peters1964}
\bibinfo{author}{\bibfnamefont{P.~C.} \bibnamefont{Peters}},
  \bibinfo{journal}{Phys. Rev.} \textbf{\bibinfo{volume}{136}},
  \bibinfo{pages}{B1224} (\bibinfo{year}{1964}).

\bibitem[{\citenamefont{Ohme}(2012)}]{Ohme:2011rm}
\bibinfo{author}{\bibfnamefont{F.}~\bibnamefont{Ohme}},
  \bibinfo{journal}{Class.Quant.Grav.} \textbf{\bibinfo{volume}{29}},
  \bibinfo{pages}{124002} (\bibinfo{year}{2012}), \eprint{1111.3737}.

\bibitem[{\citenamefont{Baker et~al.}(2007{\natexlab{a}})\citenamefont{Baker,
  McWilliams, van Meter, Centrella, Choi, Kelly, and Koppitz}}]{Baker2006e}
\bibinfo{author}{\bibfnamefont{J.~G.} \bibnamefont{Baker}},
  \bibinfo{author}{\bibfnamefont{S.~T.} \bibnamefont{McWilliams}},
  \bibinfo{author}{\bibfnamefont{J.~R.} \bibnamefont{van Meter}},
  \bibinfo{author}{\bibfnamefont{J.}~\bibnamefont{Centrella}},
  \bibinfo{author}{\bibfnamefont{D.-I.} \bibnamefont{Choi}},
  \bibinfo{author}{\bibfnamefont{B.~J.} \bibnamefont{Kelly}}, \bibnamefont{and}
  \bibinfo{author}{\bibfnamefont{M.}~\bibnamefont{Koppitz}},
  \bibinfo{journal}{Phys.\ Rev.\ D} \textbf{\bibinfo{volume}{75}},
  \bibinfo{pages}{124024} (\bibinfo{year}{2007}{\natexlab{a}}).

\bibitem[{\citenamefont{Br{\"u}gmann et~al.}(2008)\citenamefont{Br{\"u}gmann,
  Gonz\'{a}lez, Hannam, Husa, Sperhake, and Tichy}}]{Bruegmann2006}
\bibinfo{author}{\bibfnamefont{B.}~\bibnamefont{Br{\"u}gmann}},
  \bibinfo{author}{\bibfnamefont{J.~A.} \bibnamefont{Gonz\'{a}lez}},
  \bibinfo{author}{\bibfnamefont{M.}~\bibnamefont{Hannam}},
  \bibinfo{author}{\bibfnamefont{S.}~\bibnamefont{Husa}},
  \bibinfo{author}{\bibfnamefont{U.}~\bibnamefont{Sperhake}}, \bibnamefont{and}
  \bibinfo{author}{\bibfnamefont{W.}~\bibnamefont{Tichy}},
  \bibinfo{journal}{Phys.\ Rev.\ D} \textbf{\bibinfo{volume}{77}},
  \bibinfo{eid}{024027} (\bibinfo{year}{2008}).

\bibitem[{\citenamefont{Campanelli
  et~al.}(2006{\natexlab{a}})\citenamefont{Campanelli, Lousto, Marronetti, and
  Zlochower}}]{Campanelli2006a}
\bibinfo{author}{\bibfnamefont{M.}~\bibnamefont{Campanelli}},
  \bibinfo{author}{\bibfnamefont{C.}~\bibnamefont{Lousto}},
  \bibinfo{author}{\bibfnamefont{P.}~\bibnamefont{Marronetti}},
  \bibnamefont{and}
  \bibinfo{author}{\bibfnamefont{Y.}~\bibnamefont{Zlochower}},
  \bibinfo{journal}{Phys.Rev.Lett.} \textbf{\bibinfo{volume}{96}},
  \bibinfo{pages}{111101} (\bibinfo{year}{2006}{\natexlab{a}}),
  \eprint{gr-qc/0511048}.

\bibitem[{\citenamefont{Campanelli
  et~al.}(2006{\natexlab{b}})\citenamefont{Campanelli, Lousto, and
  Zlochower}}]{Campanelli2006c}
\bibinfo{author}{\bibfnamefont{M.}~\bibnamefont{Campanelli}},
  \bibinfo{author}{\bibfnamefont{C.~O.} \bibnamefont{Lousto}},
  \bibnamefont{and}
  \bibinfo{author}{\bibfnamefont{Y.}~\bibnamefont{Zlochower}},
  \bibinfo{journal}{Phys.\ Rev.\ D} \textbf{\bibinfo{volume}{74}},
  \bibinfo{pages}{041501(R)} (\bibinfo{year}{2006}{\natexlab{b}}),
  \eprint{gr-qc/0604012}.

\bibitem[{\citenamefont{Husa et~al.}(2008{\natexlab{a}})\citenamefont{Husa,
  Gonz{\'a}lez, Hannam, Br{\"u}gmann, and Sperhake}}]{Husa2007}
\bibinfo{author}{\bibfnamefont{S.}~\bibnamefont{Husa}},
  \bibinfo{author}{\bibfnamefont{J.~A.} \bibnamefont{Gonz{\'a}lez}},
  \bibinfo{author}{\bibfnamefont{M.}~\bibnamefont{Hannam}},
  \bibinfo{author}{\bibfnamefont{B.}~\bibnamefont{Br{\"u}gmann}},
  \bibnamefont{and} \bibinfo{author}{\bibfnamefont{U.}~\bibnamefont{Sperhake}},
  \bibinfo{journal}{Class.\ Quantum Grav.} \textbf{\bibinfo{volume}{25}},
  \bibinfo{pages}{105006} (\bibinfo{year}{2008}{\natexlab{a}}).

\bibitem[{\citenamefont{Boyle et~al.}(2008{\natexlab{a}})\citenamefont{Boyle,
  Kesden, and Nissanke}}]{Boyle2007a}
\bibinfo{author}{\bibfnamefont{L.}~\bibnamefont{Boyle}},
  \bibinfo{author}{\bibfnamefont{M.}~\bibnamefont{Kesden}}, \bibnamefont{and}
  \bibinfo{author}{\bibfnamefont{S.}~\bibnamefont{Nissanke}},
  \bibinfo{journal}{Phys. Rev. Lett.} \textbf{\bibinfo{volume}{100}},
  \bibinfo{pages}{151101} (\bibinfo{year}{2008}{\natexlab{a}}),
  \bibinfo{note}{arXiv:0709.0299 [gr-qc]}.

\bibitem[{\citenamefont{Baker et~al.}(2007{\natexlab{b}})\citenamefont{Baker,
  Campanelli, Pretorius, and Zlochower}}]{Baker-Campanelli-etal:2007}
\bibinfo{author}{\bibfnamefont{J.~G.} \bibnamefont{Baker}},
  \bibinfo{author}{\bibfnamefont{M.}~\bibnamefont{Campanelli}},
  \bibinfo{author}{\bibfnamefont{F.}~\bibnamefont{Pretorius}},
  \bibnamefont{and}
  \bibinfo{author}{\bibfnamefont{Y.}~\bibnamefont{Zlochower}},
  \bibinfo{journal}{Class.\ Quantum Grav.} \textbf{\bibinfo{volume}{24}},
  \bibinfo{pages}{S25} (\bibinfo{year}{2007}{\natexlab{b}}).

\bibitem[{\citenamefont{Bode et~al.}(2008)\citenamefont{Bode, Shoemaker,
  Herrmann, and Hinder}}]{BodeEtAl:2008}
\bibinfo{author}{\bibfnamefont{T.}~\bibnamefont{Bode}},
  \bibinfo{author}{\bibfnamefont{D.}~\bibnamefont{Shoemaker}},
  \bibinfo{author}{\bibfnamefont{F.}~\bibnamefont{Herrmann}}, \bibnamefont{and}
  \bibinfo{author}{\bibfnamefont{I.}~\bibnamefont{Hinder}},
  \bibinfo{journal}{Phys.\ Rev.\ D} \textbf{\bibinfo{volume}{77}},
  \bibinfo{pages}{44027} (\bibinfo{year}{2008}).

\bibitem[{\citenamefont{Chu et~al.}(2009)\citenamefont{Chu, Pfeiffer, and
  Scheel}}]{Chu2009}
\bibinfo{author}{\bibfnamefont{T.}~\bibnamefont{Chu}},
  \bibinfo{author}{\bibfnamefont{H.~P.} \bibnamefont{Pfeiffer}},
  \bibnamefont{and} \bibinfo{author}{\bibfnamefont{M.~A.}
  \bibnamefont{Scheel}}, \bibinfo{journal}{Phys.\ Rev.\ D}
  \textbf{\bibinfo{volume}{80}}, \bibinfo{pages}{124051}
  (\bibinfo{year}{2009}), \eprint{0909.1313}.

\bibitem[{\citenamefont{{M. A. Scheel, M. Boyle, T. Chu, L. E. Kidder, K. D.
  Matthews and H. P. Pfeiffer}}(2009)}]{Scheel2009}
\bibinfo{author}{\bibnamefont{{M. A. Scheel, M. Boyle, T. Chu, L. E. Kidder, K.
  D. Matthews and H. P. Pfeiffer}}}, \bibinfo{journal}{Phys.\ Rev.\ D}
  \textbf{\bibinfo{volume}{79}}, \bibinfo{pages}{024003}
  (\bibinfo{year}{2009}), \eprint{arXiv:gr-qc/0810.1767}.

\bibitem[{\citenamefont{Hannam et~al.}(2009)\citenamefont{Hannam, Husa, Baker,
  Boyle, Bruegmann, Chu, Dorband, Herrmann, Hinder, Kelly
  et~al.}}]{Hannam:2009hh}
\bibinfo{author}{\bibfnamefont{M.}~\bibnamefont{Hannam}},
  \bibinfo{author}{\bibfnamefont{S.}~\bibnamefont{Husa}},
  \bibinfo{author}{\bibfnamefont{J.~G.} \bibnamefont{Baker}},
  \bibinfo{author}{\bibfnamefont{M.}~\bibnamefont{Boyle}},
  \bibinfo{author}{\bibfnamefont{B.}~\bibnamefont{Bruegmann}},
  \bibinfo{author}{\bibfnamefont{T.}~\bibnamefont{Chu}},
  \bibinfo{author}{\bibfnamefont{N.}~\bibnamefont{Dorband}},
  \bibinfo{author}{\bibfnamefont{F.}~\bibnamefont{Herrmann}},
  \bibinfo{author}{\bibfnamefont{I.}~\bibnamefont{Hinder}},
  \bibinfo{author}{\bibfnamefont{B.~J.} \bibnamefont{Kelly}},
  \bibnamefont{et~al.}, \bibinfo{journal}{Phys.\ Rev.\ D}
  \textbf{\bibinfo{volume}{79}}, \bibinfo{pages}{084025}
  (\bibinfo{year}{2009}), \eprint{arXiv:0901.2437}.

\bibitem[{\citenamefont{Lovelace et~al.}(2011)\citenamefont{Lovelace, Scheel,
  and Szilagyi}}]{Lovelace:2010ne}
\bibinfo{author}{\bibfnamefont{G.}~\bibnamefont{Lovelace}},
  \bibinfo{author}{\bibfnamefont{M.}~\bibnamefont{Scheel}}, \bibnamefont{and}
  \bibinfo{author}{\bibfnamefont{B.}~\bibnamefont{Szilagyi}},
  \bibinfo{journal}{Phys.\ Rev.\ D} \textbf{\bibinfo{volume}{83}},
  \bibinfo{pages}{024010} (\bibinfo{year}{2011}), \eprint{1010.2777}.

\bibitem[{\citenamefont{Hannam et~al.}(2010)\citenamefont{Hannam, Husa, Ohme,
  M\"{u}ller, and Br\"{u}gmann}}]{HannamEtAl:2010}
\bibinfo{author}{\bibfnamefont{M.}~\bibnamefont{Hannam}},
  \bibinfo{author}{\bibfnamefont{S.}~\bibnamefont{Husa}},
  \bibinfo{author}{\bibfnamefont{F.}~\bibnamefont{Ohme}},
  \bibinfo{author}{\bibfnamefont{D.}~\bibnamefont{M\"{u}ller}},
  \bibnamefont{and}
  \bibinfo{author}{\bibfnamefont{B.}~\bibnamefont{Br\"{u}gmann}},
  \bibinfo{journal}{Phys.\ Rev.\ D} \textbf{\bibinfo{volume}{82}},
  \bibinfo{pages}{124008} (\bibinfo{year}{2010}), \eprint{arXiv:1007.4789}.

\bibitem[{\citenamefont{Lovelace et~al.}(2012)\citenamefont{Lovelace, Boyle,
  Scheel, and Szil\'{a}gyi}}]{Lovelace:2011nu}
\bibinfo{author}{\bibfnamefont{G.}~\bibnamefont{Lovelace}},
  \bibinfo{author}{\bibfnamefont{M.}~\bibnamefont{Boyle}},
  \bibinfo{author}{\bibfnamefont{M.~A.} \bibnamefont{Scheel}},
  \bibnamefont{and}
  \bibinfo{author}{\bibfnamefont{B.}~\bibnamefont{Szil\'{a}gyi}},
  \bibinfo{journal}{Class. Quant. Grav.} \textbf{\bibinfo{volume}{29}},
  \bibinfo{pages}{045003} (\bibinfo{year}{2012}), \eprint{arXiv:1110.2229}.

\bibitem[{\citenamefont{Buchman et~al.}(2012)\citenamefont{Buchman, Pfeiffer,
  Scheel, and Szilagyi}}]{Buchman:2012dw}
\bibinfo{author}{\bibfnamefont{L.~T.} \bibnamefont{Buchman}},
  \bibinfo{author}{\bibfnamefont{H.~P.} \bibnamefont{Pfeiffer}},
  \bibinfo{author}{\bibfnamefont{M.~A.} \bibnamefont{Scheel}},
  \bibnamefont{and} \bibinfo{author}{\bibfnamefont{B.}~\bibnamefont{Szilagyi}},
  \emph{\bibinfo{title}{{Simulations of unequal mass binary black holes with
  spectral methods}}} (\bibinfo{year}{2012}), \eprint{1206.3015}.

\bibitem[{\citenamefont{{Hannam} et~al.}(2010)\citenamefont{{Hannam}, {Husa},
  {Ohme}, and {Ajith}}}]{Hannam:2010}
\bibinfo{author}{\bibfnamefont{M.}~\bibnamefont{{Hannam}}},
  \bibinfo{author}{\bibfnamefont{S.}~\bibnamefont{{Husa}}},
  \bibinfo{author}{\bibfnamefont{F.}~\bibnamefont{{Ohme}}}, \bibnamefont{and}
  \bibinfo{author}{\bibfnamefont{P.}~\bibnamefont{{Ajith}}},
  \bibinfo{journal}{Phys. Rev. D} \textbf{\bibinfo{volume}{82}},
  \bibinfo{pages}{124052} (\bibinfo{year}{2010}).

\bibitem[{\citenamefont{Ajith et~al.}(2009{\natexlab{b}})\citenamefont{Ajith,
  Babak, Chen, Hewitson, Krishnan, Sintes, Whelan, Br\"ugmann, Diener, Dorband
  et~al.}}]{Ajith:2008b}
\bibinfo{author}{\bibfnamefont{P.}~\bibnamefont{Ajith}},
  \bibinfo{author}{\bibfnamefont{S.}~\bibnamefont{Babak}},
  \bibinfo{author}{\bibfnamefont{Y.}~\bibnamefont{Chen}},
  \bibinfo{author}{\bibfnamefont{M.}~\bibnamefont{Hewitson}},
  \bibinfo{author}{\bibfnamefont{B.}~\bibnamefont{Krishnan}},
  \bibinfo{author}{\bibfnamefont{A.~M.} \bibnamefont{Sintes}},
  \bibinfo{author}{\bibfnamefont{J.~T.} \bibnamefont{Whelan}},
  \bibinfo{author}{\bibfnamefont{B.}~\bibnamefont{Br\"ugmann}},
  \bibinfo{author}{\bibfnamefont{P.}~\bibnamefont{Diener}},
  \bibinfo{author}{\bibfnamefont{N.}~\bibnamefont{Dorband}},
  \bibnamefont{et~al.}, \bibinfo{journal}{Phys. Rev. D}
  \textbf{\bibinfo{volume}{79}}, \bibinfo{pages}{129901}
  (\bibinfo{year}{2009}{\natexlab{b}}).

\bibitem[{\citenamefont{{Ajith} et~al.}(2009)\citenamefont{{Ajith}, {Hannam},
  {Husa}, {Chen}, {Bruegmann}, {Dorband}, {Mueller}, {Ohme}, {Pollney},
  {Reisswig} et~al.}}]{Ajith:2009}
\bibinfo{author}{\bibfnamefont{P.}~\bibnamefont{{Ajith}}},
  \bibinfo{author}{\bibfnamefont{M.}~\bibnamefont{{Hannam}}},
  \bibinfo{author}{\bibfnamefont{S.}~\bibnamefont{{Husa}}},
  \bibinfo{author}{\bibfnamefont{Y.}~\bibnamefont{{Chen}}},
  \bibinfo{author}{\bibfnamefont{B.}~\bibnamefont{{Bruegmann}}},
  \bibinfo{author}{\bibfnamefont{N.}~\bibnamefont{{Dorband}}},
  \bibinfo{author}{\bibfnamefont{D.}~\bibnamefont{{Mueller}}},
  \bibinfo{author}{\bibfnamefont{F.}~\bibnamefont{{Ohme}}},
  \bibinfo{author}{\bibfnamefont{D.}~\bibnamefont{{Pollney}}},
  \bibinfo{author}{\bibfnamefont{C.}~\bibnamefont{{Reisswig}}},
  \bibnamefont{et~al.}, \bibinfo{journal}{ArXiv e-prints}
  (\bibinfo{year}{2009}), \eprint{0909.2867}.

\bibitem[{\citenamefont{Tiec et~al.}(2011)\citenamefont{Tiec, Mrou\'{e},
  Barack, Buonanno, Pfeiffer et~al.}}]{Tiec:2011bk}
\bibinfo{author}{\bibfnamefont{A.~L.} \bibnamefont{Tiec}},
  \bibinfo{author}{\bibfnamefont{A.~H.} \bibnamefont{Mrou\'{e}}},
  \bibinfo{author}{\bibfnamefont{L.}~\bibnamefont{Barack}},
  \bibinfo{author}{\bibfnamefont{A.}~\bibnamefont{Buonanno}},
  \bibinfo{author}{\bibfnamefont{H.~P.} \bibnamefont{Pfeiffer}},
  \bibnamefont{et~al.}, \bibinfo{journal}{Phys.Rev.Lett.}
  \textbf{\bibinfo{volume}{107}}, \bibinfo{pages}{141101}
  (\bibinfo{year}{2011}), \eprint{1106.3278}.

\bibitem[{\citenamefont{Mrou\'e et~al.}(2008)\citenamefont{Mrou\'e, Kidder, and
  Teukolsky}}]{Mroue2008}
\bibinfo{author}{\bibfnamefont{A.~H.} \bibnamefont{Mrou\'e}},
  \bibinfo{author}{\bibfnamefont{L.~E.} \bibnamefont{Kidder}},
  \bibnamefont{and} \bibinfo{author}{\bibfnamefont{S.~A.}
  \bibnamefont{Teukolsky}}, \bibinfo{journal}{Phys.\ Rev.\ D}
  \textbf{\bibinfo{volume}{78}}, \bibinfo{pages}{044004}
  (\bibinfo{year}{2008}).

\bibitem[{\citenamefont{Taracchini et~al.}(2012)\citenamefont{Taracchini,
  Buonanno, Barausse, Boyle, Chu, Lovelace, Pfeiffer, and
  Scheel}}]{Taracchini:2012}
\bibinfo{author}{\bibfnamefont{A.}~\bibnamefont{Taracchini}},
  \bibinfo{author}{\bibfnamefont{A.}~\bibnamefont{Buonanno}},
  \bibinfo{author}{\bibfnamefont{E.}~\bibnamefont{Barausse}},
  \bibinfo{author}{\bibfnamefont{M.}~\bibnamefont{Boyle}},
  \bibinfo{author}{\bibfnamefont{T.}~\bibnamefont{Chu}},
  \bibinfo{author}{\bibfnamefont{G.}~\bibnamefont{Lovelace}},
  \bibinfo{author}{\bibfnamefont{H.~P.} \bibnamefont{Pfeiffer}},
  \bibnamefont{and} \bibinfo{author}{\bibfnamefont{M.~A.} \bibnamefont{Scheel}}
  (\bibinfo{year}{2012}), \eprint{1202.0790}.

\bibitem[{\citenamefont{Boyle et~al.}(2008{\natexlab{b}})\citenamefont{Boyle,
  Buonanno, Kidder, Mrou\'{e}, Pan et~al.}}]{Boyle:2008}
\bibinfo{author}{\bibfnamefont{M.}~\bibnamefont{Boyle}},
  \bibinfo{author}{\bibfnamefont{A.}~\bibnamefont{Buonanno}},
  \bibinfo{author}{\bibfnamefont{L.~E.} \bibnamefont{Kidder}},
  \bibinfo{author}{\bibfnamefont{A.~H.} \bibnamefont{Mrou\'{e}}},
  \bibinfo{author}{\bibfnamefont{Y.}~\bibnamefont{Pan}}, \bibnamefont{et~al.},
  \bibinfo{journal}{Phys.\ Rev.\ D} \textbf{\bibinfo{volume}{78}},
  \bibinfo{pages}{104020} (\bibinfo{year}{2008}{\natexlab{b}}),
  \eprint{0804.4184}.

\bibitem[{\citenamefont{{Pan} et~al.}(2011)\citenamefont{{Pan}, {Buonanno},
  {Boyle}, {Buchman}, {Kidder}, {Pfeiffer}, and {Scheel}}}]{PanEtAl:2011}
\bibinfo{author}{\bibfnamefont{Y.}~\bibnamefont{{Pan}}},
  \bibinfo{author}{\bibfnamefont{A.}~\bibnamefont{{Buonanno}}},
  \bibinfo{author}{\bibfnamefont{M.}~\bibnamefont{{Boyle}}},
  \bibinfo{author}{\bibfnamefont{L.~T.} \bibnamefont{{Buchman}}},
  \bibinfo{author}{\bibfnamefont{L.~E.} \bibnamefont{{Kidder}}},
  \bibinfo{author}{\bibfnamefont{H.~P.} \bibnamefont{{Pfeiffer}}},
  \bibnamefont{and} \bibinfo{author}{\bibfnamefont{M.~A.}
  \bibnamefont{{Scheel}}}, \bibinfo{journal}{Phys.\ Rev.\ D}
  \textbf{\bibinfo{volume}{84}}, \bibinfo{pages}{124052}
  (\bibinfo{year}{2011}), \eprint{1106.1021}.

\bibitem[{\citenamefont{Campanelli
  et~al.}(2007{\natexlab{a}})\citenamefont{Campanelli, Lousto, Zlochower, and
  Merritt}}]{Campanelli2007a}
\bibinfo{author}{\bibfnamefont{M.}~\bibnamefont{Campanelli}},
  \bibinfo{author}{\bibfnamefont{C.~O.} \bibnamefont{Lousto}},
  \bibinfo{author}{\bibfnamefont{Y.}~\bibnamefont{Zlochower}},
  \bibnamefont{and} \bibinfo{author}{\bibfnamefont{D.}~\bibnamefont{Merritt}},
  \bibinfo{journal}{Astrophys.\ J.\ Lett.} \textbf{\bibinfo{volume}{659}},
  \bibinfo{pages}{L5} (\bibinfo{year}{2007}{\natexlab{a}}).

\bibitem[{\citenamefont{Campanelli
  et~al.}(2007{\natexlab{b}})\citenamefont{Campanelli, Lousto, Zlochower,
  Krishnan, and Merritt}}]{Campanelli2007b}
\bibinfo{author}{\bibfnamefont{M.}~\bibnamefont{Campanelli}},
  \bibinfo{author}{\bibfnamefont{C.~O.} \bibnamefont{Lousto}},
  \bibinfo{author}{\bibfnamefont{Y.}~\bibnamefont{Zlochower}},
  \bibinfo{author}{\bibfnamefont{B.}~\bibnamefont{Krishnan}}, \bibnamefont{and}
  \bibinfo{author}{\bibfnamefont{D.}~\bibnamefont{Merritt}},
  \bibinfo{journal}{Phys.\ Rev.\ D} \textbf{\bibinfo{volume}{75}},
  \bibinfo{pages}{064030} (\bibinfo{year}{2007}{\natexlab{b}}),
  \eprint{gr-qc/0612076}.

\bibitem[{\citenamefont{Schmidt et~al.}(2011)\citenamefont{Schmidt, Hannam,
  Husa, and Ajith}}]{Schmidt2010}
\bibinfo{author}{\bibfnamefont{P.}~\bibnamefont{Schmidt}},
  \bibinfo{author}{\bibfnamefont{M.}~\bibnamefont{Hannam}},
  \bibinfo{author}{\bibfnamefont{S.}~\bibnamefont{Husa}}, \bibnamefont{and}
  \bibinfo{author}{\bibfnamefont{P.}~\bibnamefont{Ajith}},
  \bibinfo{journal}{Phys.\ Rev.\ D} \textbf{\bibinfo{volume}{84}},
  \bibinfo{pages}{024046} (\bibinfo{year}{2011}), \eprint{arxiv:1012.2879}.

\bibitem[{\citenamefont{{{OShaughnessy}, R. and Vaishnav, B. and Healy, J. and
  Meeks, Z. and Shoemaker, D.}}(2011)}]{OShaughnessy2011}
\bibinfo{author}{\bibnamefont{{{OShaughnessy}, R. and Vaishnav, B. and Healy,
  J. and Meeks, Z. and Shoemaker, D.}}}, \emph{\bibinfo{title}{Efficient
  asymptotic frame selection for binary black hole spacetimes using asymptotic
  radiation}} (\bibinfo{year}{2011}), \eprint{arXiv:1109.5224}.

\bibitem[{\citenamefont{Boyle et~al.}(2011)\citenamefont{Boyle, Owen, and
  Pfeiffer}}]{Boyle:2011gg}
\bibinfo{author}{\bibfnamefont{M.}~\bibnamefont{Boyle}},
  \bibinfo{author}{\bibfnamefont{R.}~\bibnamefont{Owen}}, \bibnamefont{and}
  \bibinfo{author}{\bibfnamefont{H.~P.} \bibnamefont{Pfeiffer}},
  \bibinfo{journal}{Phys.\ Rev.\ D} \textbf{\bibinfo{volume}{84}},
  \bibinfo{pages}{124011} (\bibinfo{year}{2011}), \eprint{arXiv:1110.2965}.

\bibitem[{\citenamefont{Sturani
  et~al.}(2010{\natexlab{a}})\citenamefont{Sturani, Fischetti, Cadonati, Guidi,
  Healy et~al.}}]{Sturani:2010ju}
\bibinfo{author}{\bibfnamefont{R.}~\bibnamefont{Sturani}},
  \bibinfo{author}{\bibfnamefont{S.}~\bibnamefont{Fischetti}},
  \bibinfo{author}{\bibfnamefont{L.}~\bibnamefont{Cadonati}},
  \bibinfo{author}{\bibfnamefont{G.}~\bibnamefont{Guidi}},
  \bibinfo{author}{\bibfnamefont{J.}~\bibnamefont{Healy}}, \bibnamefont{et~al.}
  (\bibinfo{year}{2010}{\natexlab{a}}), \eprint{1012.5172}.

\bibitem[{\citenamefont{Sturani
  et~al.}(2010{\natexlab{b}})\citenamefont{Sturani, Fischetti, Cadonati, Guidi,
  Healy et~al.}}]{Sturani:2010yv}
\bibinfo{author}{\bibfnamefont{R.}~\bibnamefont{Sturani}},
  \bibinfo{author}{\bibfnamefont{S.}~\bibnamefont{Fischetti}},
  \bibinfo{author}{\bibfnamefont{L.}~\bibnamefont{Cadonati}},
  \bibinfo{author}{\bibfnamefont{G.}~\bibnamefont{Guidi}},
  \bibinfo{author}{\bibfnamefont{J.}~\bibnamefont{Healy}},
  \bibnamefont{et~al.}, \bibinfo{journal}{J.Phys.Conf.Ser.}
  \textbf{\bibinfo{volume}{243}}, \bibinfo{pages}{012007}
  (\bibinfo{year}{2010}{\natexlab{b}}), \eprint{1005.0551}.

\bibitem[{\citenamefont{O'Shaughnessy et~al.}(2012)\citenamefont{O'Shaughnessy,
  Healy, London, Meeks, and Shoemaker}}]{O'Shaughnessy:2012vm}
\bibinfo{author}{\bibfnamefont{R.}~\bibnamefont{O'Shaughnessy}},
  \bibinfo{author}{\bibfnamefont{J.}~\bibnamefont{Healy}},
  \bibinfo{author}{\bibfnamefont{L.}~\bibnamefont{London}},
  \bibinfo{author}{\bibfnamefont{Z.}~\bibnamefont{Meeks}}, \bibnamefont{and}
  \bibinfo{author}{\bibfnamefont{D.}~\bibnamefont{Shoemaker}},
  \bibinfo{journal}{Phys.Rev.} \textbf{\bibinfo{volume}{D85}},
  \bibinfo{pages}{084003} (\bibinfo{year}{2012}), \eprint{1201.2113}.

\bibitem[{\citenamefont{Buonanno
  et~al.}(2007{\natexlab{b}})\citenamefont{Buonanno, Cook, and
  Pretorius}}]{Buonanno-Cook-Pretorius:2007}
\bibinfo{author}{\bibfnamefont{A.}~\bibnamefont{Buonanno}},
  \bibinfo{author}{\bibfnamefont{G.~B.} \bibnamefont{Cook}}, \bibnamefont{and}
  \bibinfo{author}{\bibfnamefont{F.}~\bibnamefont{Pretorius}},
  \bibinfo{journal}{Phys.\ Rev.\ D} \textbf{\bibinfo{volume}{75}},
  \bibinfo{pages}{124018} (\bibinfo{year}{2007}{\natexlab{b}}),
  \eprint{gr-qc/0610122}.

\bibitem[{\citenamefont{Pfeiffer et~al.}(2007)\citenamefont{Pfeiffer, Brown,
  Kidder, Lindblom, Lovelace, and Scheel}}]{Pfeiffer-Brown-etal:2007}
\bibinfo{author}{\bibfnamefont{H.~P.} \bibnamefont{Pfeiffer}},
  \bibinfo{author}{\bibfnamefont{D.~A.} \bibnamefont{Brown}},
  \bibinfo{author}{\bibfnamefont{L.~E.} \bibnamefont{Kidder}},
  \bibinfo{author}{\bibfnamefont{L.}~\bibnamefont{Lindblom}},
  \bibinfo{author}{\bibfnamefont{G.}~\bibnamefont{Lovelace}}, \bibnamefont{and}
  \bibinfo{author}{\bibfnamefont{M.~A.} \bibnamefont{Scheel}},
  \bibinfo{journal}{Class.\ Quantum Grav.} \textbf{\bibinfo{volume}{24}},
  \bibinfo{pages}{S59} (\bibinfo{year}{2007}).

\bibitem[{\citenamefont{Baker et~al.}(2007{\natexlab{c}})\citenamefont{Baker,
  van Meter, McWilliams, Centrella, and Kelly}}]{Baker2006d}
\bibinfo{author}{\bibfnamefont{J.~G.} \bibnamefont{Baker}},
  \bibinfo{author}{\bibfnamefont{J.~R.} \bibnamefont{van Meter}},
  \bibinfo{author}{\bibfnamefont{S.~T.} \bibnamefont{McWilliams}},
  \bibinfo{author}{\bibfnamefont{J.}~\bibnamefont{Centrella}},
  \bibnamefont{and} \bibinfo{author}{\bibfnamefont{B.~J.} \bibnamefont{Kelly}},
  \bibinfo{journal}{Phys.\ Rev.\ Lett.} \textbf{\bibinfo{volume}{99}},
  \bibinfo{eid}{181101} (\bibinfo{year}{2007}{\natexlab{c}}).

\bibitem[{\citenamefont{Lincoln and Will}(1990)}]{Lincoln-Will:1990}
\bibinfo{author}{\bibfnamefont{C.}~\bibnamefont{Lincoln}} \bibnamefont{and}
  \bibinfo{author}{\bibfnamefont{C.}~\bibnamefont{Will}},
  \bibinfo{journal}{Phys.\ Rev.\ D} \textbf{\bibinfo{volume}{42}},
  \bibinfo{pages}{1123} (\bibinfo{year}{1990}).

\bibitem[{\citenamefont{Damour and Sch\"afer}(1988)}]{Damour-Schafer:1988}
\bibinfo{author}{\bibfnamefont{T.}~\bibnamefont{Damour}} \bibnamefont{and}
  \bibinfo{author}{\bibfnamefont{G.}~\bibnamefont{Sch\"afer}},
  \bibinfo{journal}{Nuovo Cimento Soc. Ital. Fis.} \textbf{\bibinfo{volume}{101
  B}}, \bibinfo{pages}{127} (\bibinfo{year}{1988}).

\bibitem[{\citenamefont{Damour et~al.}(2004)\citenamefont{Damour, Gopakumar,
  and Iyer}}]{Damour2004}
\bibinfo{author}{\bibfnamefont{T.}~\bibnamefont{Damour}},
  \bibinfo{author}{\bibfnamefont{A.}~\bibnamefont{Gopakumar}},
  \bibnamefont{and} \bibinfo{author}{\bibfnamefont{B.~R.} \bibnamefont{Iyer}},
  \bibinfo{journal}{Phys.\ Rev.\ D} \textbf{\bibinfo{volume}{70}},
  \bibinfo{pages}{064028} (\bibinfo{year}{2004}).

\bibitem[{\citenamefont{K\"{o}nigsd\"{o}rffer and
  Gopakumar}(2006)}]{KonigsdorfferGopakumar2006}
\bibinfo{author}{\bibfnamefont{C.}~\bibnamefont{K\"{o}nigsd\"{o}rffer}}
  \bibnamefont{and}
  \bibinfo{author}{\bibfnamefont{A.}~\bibnamefont{Gopakumar}},
  \bibinfo{journal}{Phys.\ Rev.\ D} \textbf{\bibinfo{volume}{73}},
  \bibinfo{eid}{124012} (\bibinfo{year}{2006}), \eprint{gr-qc/0603056}.

\bibitem[{\citenamefont{{R.M. Memmesheimer, A. Gopakumar and G.
  Sch\"afer}}(2004)}]{Memmesheimer-etal:2004}
\bibinfo{author}{\bibnamefont{{R.M. Memmesheimer, A. Gopakumar and G.
  Sch\"afer}}}, \bibinfo{journal}{Phys.\ Rev.\ D}
  \textbf{\bibinfo{volume}{70}}, \bibinfo{pages}{104011}
  (\bibinfo{year}{2004}).

\bibitem[{\citenamefont{Berti et~al.}(2006)\citenamefont{Berti, Iyer, and
  Will}}]{Berti2006}
\bibinfo{author}{\bibfnamefont{E.}~\bibnamefont{Berti}},
  \bibinfo{author}{\bibfnamefont{S.}~\bibnamefont{Iyer}}, \bibnamefont{and}
  \bibinfo{author}{\bibfnamefont{C.~M.} \bibnamefont{Will}},
  \bibinfo{journal}{Phys.\ Rev.\ D} \textbf{\bibinfo{volume}{74}},
  \bibinfo{eid}{061503} (\bibinfo{year}{2006}), \eprint{gr-qc/0607047}.

\bibitem[{\citenamefont{Mora and Will}(2002)}]{Will-Mora:2002}
\bibinfo{author}{\bibfnamefont{T.}~\bibnamefont{Mora}} \bibnamefont{and}
  \bibinfo{author}{\bibfnamefont{C.}~\bibnamefont{Will}},
  \bibinfo{journal}{Phys.\ Rev.\ D} \textbf{\bibinfo{volume}{66}},
  \bibinfo{pages}{101501(R)} (\bibinfo{year}{2002}).

\bibitem[{\citenamefont{Husa et~al.}(2008{\natexlab{b}})\citenamefont{Husa,
  Hannam, Gonz{\'a}lez, Sperhake, and Br{\"u}gmann}}]{Husa-Hannam-etal:2007}
\bibinfo{author}{\bibfnamefont{S.}~\bibnamefont{Husa}},
  \bibinfo{author}{\bibfnamefont{M.}~\bibnamefont{Hannam}},
  \bibinfo{author}{\bibfnamefont{J.~A.} \bibnamefont{Gonz{\'a}lez}},
  \bibinfo{author}{\bibfnamefont{U.}~\bibnamefont{Sperhake}}, \bibnamefont{and}
  \bibinfo{author}{\bibfnamefont{B.}~\bibnamefont{Br{\"u}gmann}},
  \bibinfo{journal}{Phys.\ Rev.\ D} \textbf{\bibinfo{volume}{77}},
  \bibinfo{pages}{044037} (\bibinfo{year}{2008}{\natexlab{b}}),
  \eprint{0706.0904}.

\bibitem[{\citenamefont{Campanelli et~al.}(2009)\citenamefont{Campanelli,
  Lousto, Nakano, and Zlochower}}]{CampanelliEtal2009}
\bibinfo{author}{\bibfnamefont{M.}~\bibnamefont{Campanelli}},
  \bibinfo{author}{\bibfnamefont{C.~O.} \bibnamefont{Lousto}},
  \bibinfo{author}{\bibfnamefont{H.}~\bibnamefont{Nakano}}, \bibnamefont{and}
  \bibinfo{author}{\bibfnamefont{Y.}~\bibnamefont{Zlochower}},
  \bibinfo{journal}{Phys.\ Rev.\ D} \textbf{\bibinfo{volume}{79}},
  \bibinfo{pages}{84010} (\bibinfo{year}{2009}),
  \eprint{arXiv:gr-qc/0808.0713}.

\bibitem[{\citenamefont{Mrou\'{e} et~al.}(2010)\citenamefont{Mrou\'{e},
  Pfeiffer, Kidder, and Teukolsky}}]{Mroue2010}
\bibinfo{author}{\bibfnamefont{A.~H.} \bibnamefont{Mrou\'{e}}},
  \bibinfo{author}{\bibfnamefont{H.~P.} \bibnamefont{Pfeiffer}},
  \bibinfo{author}{\bibfnamefont{L.~E.} \bibnamefont{Kidder}},
  \bibnamefont{and} \bibinfo{author}{\bibfnamefont{S.~A.}
  \bibnamefont{Teukolsky}}, \bibinfo{journal}{Phys.\ Rev.\ D}
  \textbf{\bibinfo{volume}{82}}, \bibinfo{pages}{124016}
  (\bibinfo{year}{2010}), \eprint{arXiv:1004.4697 [gr-qc]}.

\bibitem[{\citenamefont{Buonanno et~al.}(2011)\citenamefont{Buonanno, Kidder,
  Mrou\'{e}, Pfeiffer, and Taracchini}}]{Buonanno:2010yk}
\bibinfo{author}{\bibfnamefont{A.}~\bibnamefont{Buonanno}},
  \bibinfo{author}{\bibfnamefont{L.~E.} \bibnamefont{Kidder}},
  \bibinfo{author}{\bibfnamefont{A.~H.} \bibnamefont{Mrou\'{e}}},
  \bibinfo{author}{\bibfnamefont{H.~P.} \bibnamefont{Pfeiffer}},
  \bibnamefont{and}
  \bibinfo{author}{\bibfnamefont{A.}~\bibnamefont{Taracchini}},
  \bibinfo{journal}{Phys.Rev.} \textbf{\bibinfo{volume}{D83}},
  \bibinfo{pages}{104034} (\bibinfo{year}{2011}), \eprint{1012.1549}.

\bibitem[{\citenamefont{Boyle et~al.}(2007)\citenamefont{Boyle, Brown, Kidder,
  Mrou{\'e}, Pfeiffer, Scheel, Cook, and Teukolsky}}]{Boyle2007}
\bibinfo{author}{\bibfnamefont{M.}~\bibnamefont{Boyle}},
  \bibinfo{author}{\bibfnamefont{D.~A.} \bibnamefont{Brown}},
  \bibinfo{author}{\bibfnamefont{L.~E.} \bibnamefont{Kidder}},
  \bibinfo{author}{\bibfnamefont{A.~H.} \bibnamefont{Mrou{\'e}}},
  \bibinfo{author}{\bibfnamefont{H.~P.} \bibnamefont{Pfeiffer}},
  \bibinfo{author}{\bibfnamefont{M.~A.} \bibnamefont{Scheel}},
  \bibinfo{author}{\bibfnamefont{G.~B.} \bibnamefont{Cook}}, \bibnamefont{and}
  \bibinfo{author}{\bibfnamefont{S.~A.} \bibnamefont{Teukolsky}},
  \bibinfo{journal}{Phys.\ Rev.\ D} \textbf{\bibinfo{volume}{76}},
  \bibinfo{eid}{124038} (\bibinfo{year}{2007}).

\bibitem[{\citenamefont{Tichy and Marronetti}(2010)}]{Tichy:2010qa}
\bibinfo{author}{\bibfnamefont{W.}~\bibnamefont{Tichy}} \bibnamefont{and}
  \bibinfo{author}{\bibfnamefont{P.}~\bibnamefont{Marronetti}}
  (\bibinfo{year}{2010}), \eprint{1010.2936}.

\bibitem[{\citenamefont{Purrer et~al.}(2012)\citenamefont{Purrer, Husa, and
  Hannam}}]{Purrer:2012wy}
\bibinfo{author}{\bibfnamefont{M.}~\bibnamefont{Purrer}},
  \bibinfo{author}{\bibfnamefont{S.}~\bibnamefont{Husa}}, \bibnamefont{and}
  \bibinfo{author}{\bibfnamefont{M.}~\bibnamefont{Hannam}},
  \bibinfo{journal}{Phys.Rev.} \textbf{\bibinfo{volume}{D85}},
  \bibinfo{pages}{124051} (\bibinfo{year}{2012}), \eprint{1203.4258}.

\bibitem[{\citenamefont{Buchman et~al.}()\citenamefont{Buchman, Pfeiffer,
  Scheel, and Szilagyi}}]{Buchman-etal-in-prep}
\bibinfo{author}{\bibfnamefont{L.~T.} \bibnamefont{Buchman}},
  \bibinfo{author}{\bibfnamefont{H.~P.} \bibnamefont{Pfeiffer}},
  \bibinfo{author}{\bibfnamefont{M.~A.} \bibnamefont{Scheel}},
  \bibnamefont{and} \bibinfo{author}{\bibfnamefont{B.}~\bibnamefont{Szilagyi}},
  \emph{\bibinfo{title}{Simulations of non-equal mass black hole binaries}},
  \bibinfo{note}{in preparation}.

\bibitem[{\citenamefont{{D. Brown and P.
  Zimmerman.}}(2010)}]{BrownZimmerman2009}
\bibinfo{author}{\bibnamefont{{D. Brown and P. Zimmerman.}}},
  \bibinfo{journal}{Phys.\ Rev.\ D} \textbf{\bibinfo{volume}{81}},
  \bibinfo{pages}{024007} (\bibinfo{year}{2010}),
  \eprint{arXiv:gr-qc/0909.0066}.

\bibitem[{\citenamefont{York}(1999)}]{York1999}
\bibinfo{author}{\bibfnamefont{J.~W.} \bibnamefont{York}},
  \bibinfo{journal}{Phys.\ Rev.\ Lett.} \textbf{\bibinfo{volume}{82}},
  \bibinfo{pages}{1350} (\bibinfo{year}{1999}).

\bibitem[{\citenamefont{Pfeiffer and York}(2003)}]{Pfeiffer2003b}
\bibinfo{author}{\bibfnamefont{H.~P.} \bibnamefont{Pfeiffer}} \bibnamefont{and}
  \bibinfo{author}{\bibfnamefont{J.~W.} \bibnamefont{York}},
  \bibinfo{journal}{Phys.\ Rev.\ D} \textbf{\bibinfo{volume}{67}},
  \bibinfo{pages}{044022} (\bibinfo{year}{2003}).

\bibitem[{\citenamefont{Cook}(2002)}]{Cook2002}
\bibinfo{author}{\bibfnamefont{G.~B.} \bibnamefont{Cook}},
  \bibinfo{journal}{Phys.\ Rev.\ D} \textbf{\bibinfo{volume}{65}},
  \bibinfo{pages}{084003} (\bibinfo{year}{2002}).

\bibitem[{\citenamefont{Cook and Pfeiffer}(2004)}]{Cook2004}
\bibinfo{author}{\bibfnamefont{G.~B.} \bibnamefont{Cook}} \bibnamefont{and}
  \bibinfo{author}{\bibfnamefont{H.~P.} \bibnamefont{Pfeiffer}},
  \bibinfo{journal}{Phys.\ Rev.\ D} \textbf{\bibinfo{volume}{70}},
  \bibinfo{pages}{104016} (\bibinfo{year}{2004}).

\bibitem[{\citenamefont{{Caudill} et~al.}(2006)\citenamefont{{Caudill}, {Cook},
  {Grigsby}, and {Pfeiffer}}}]{Caudill-etal:2006}
\bibinfo{author}{\bibfnamefont{M.}~\bibnamefont{{Caudill}}},
  \bibinfo{author}{\bibfnamefont{G.~B.} \bibnamefont{{Cook}}},
  \bibinfo{author}{\bibfnamefont{J.~D.} \bibnamefont{{Grigsby}}},
  \bibnamefont{and} \bibinfo{author}{\bibfnamefont{H.~P.}
  \bibnamefont{{Pfeiffer}}}, \bibinfo{journal}{Phys.\ Rev.\ D}
  \textbf{\bibinfo{volume}{74}}, \bibinfo{pages}{064011}
  (\bibinfo{year}{2006}).

\bibitem[{\citenamefont{Pfeiffer et~al.}(2003)\citenamefont{Pfeiffer, Kidder,
  Scheel, and Teukolsky}}]{Pfeiffer2003}
\bibinfo{author}{\bibfnamefont{H.~P.} \bibnamefont{Pfeiffer}},
  \bibinfo{author}{\bibfnamefont{L.~E.} \bibnamefont{Kidder}},
  \bibinfo{author}{\bibfnamefont{M.~A.} \bibnamefont{Scheel}},
  \bibnamefont{and} \bibinfo{author}{\bibfnamefont{S.~A.}
  \bibnamefont{Teukolsky}}, \bibinfo{journal}{Comput.\ Phys.\ Commun.}
  \textbf{\bibinfo{volume}{152}}, \bibinfo{pages}{253} (\bibinfo{year}{2003}).

\bibitem[{SpE()}]{SpECwebsite}
\bibinfo{howpublished}{\url{http://www.black-holes.org/SpEC.html}}.

\bibitem[{\citenamefont{Lindblom et~al.}(2006)\citenamefont{Lindblom, Scheel,
  Kidder, Owen, and Rinne}}]{Lindblom2006}
\bibinfo{author}{\bibfnamefont{L.}~\bibnamefont{Lindblom}},
  \bibinfo{author}{\bibfnamefont{M.~A.} \bibnamefont{Scheel}},
  \bibinfo{author}{\bibfnamefont{L.~E.} \bibnamefont{Kidder}},
  \bibinfo{author}{\bibfnamefont{R.}~\bibnamefont{Owen}}, \bibnamefont{and}
  \bibinfo{author}{\bibfnamefont{O.}~\bibnamefont{Rinne}},
  \bibinfo{journal}{Class.\ Quantum Grav.} \textbf{\bibinfo{volume}{23}},
  \bibinfo{pages}{S447} (\bibinfo{year}{2006}).

\bibitem[{\citenamefont{Friedrich}(1985)}]{Friedrich1985}
\bibinfo{author}{\bibfnamefont{H.}~\bibnamefont{Friedrich}},
  \bibinfo{journal}{Commun.\ Math.\ Phys.} \textbf{\bibinfo{volume}{100}},
  \bibinfo{pages}{525} (\bibinfo{year}{1985}).

\bibitem[{\citenamefont{Garfinkle}(2002)}]{Garfinkle2002}
\bibinfo{author}{\bibfnamefont{D.}~\bibnamefont{Garfinkle}},
  \bibinfo{journal}{Phys.\ Rev.\ D} \textbf{\bibinfo{volume}{65}},
  \bibinfo{pages}{044029} (\bibinfo{year}{2002}).

\bibitem[{\citenamefont{Gundlach et~al.}(2005)\citenamefont{Gundlach,
  Martin-Garcia, Calabrese, and Hinder}}]{Gundlach2005}
\bibinfo{author}{\bibfnamefont{C.}~\bibnamefont{Gundlach}},
  \bibinfo{author}{\bibfnamefont{J.~M.} \bibnamefont{Martin-Garcia}},
  \bibinfo{author}{\bibfnamefont{G.}~\bibnamefont{Calabrese}},
  \bibnamefont{and} \bibinfo{author}{\bibfnamefont{I.}~\bibnamefont{Hinder}},
  \bibinfo{journal}{Class.\ Quantum Grav.} \textbf{\bibinfo{volume}{22}},
  \bibinfo{pages}{3767} (\bibinfo{year}{2005}).

\bibitem[{\citenamefont{Rinne}(2006)}]{Rinne2006}
\bibinfo{author}{\bibfnamefont{O.}~\bibnamefont{Rinne}},
  \bibinfo{journal}{Class.\ Quantum Grav.} \textbf{\bibinfo{volume}{23}},
  \bibinfo{pages}{6275} (\bibinfo{year}{2006}).

\bibitem[{\citenamefont{Rinne et~al.}(2007)\citenamefont{Rinne, Lindblom, and
  Scheel}}]{Rinne2007}
\bibinfo{author}{\bibfnamefont{O.}~\bibnamefont{Rinne}},
  \bibinfo{author}{\bibfnamefont{L.}~\bibnamefont{Lindblom}}, \bibnamefont{and}
  \bibinfo{author}{\bibfnamefont{M.~A.} \bibnamefont{Scheel}},
  \bibinfo{journal}{Class.\ Quantum Grav.} \textbf{\bibinfo{volume}{24}},
  \bibinfo{pages}{4053} (\bibinfo{year}{2007}).

\bibitem[{\citenamefont{Stewart}(1998)}]{Stewart1998}
\bibinfo{author}{\bibfnamefont{J.~M.} \bibnamefont{Stewart}},
  \bibinfo{journal}{Class.\ Quantum Grav.} \textbf{\bibinfo{volume}{15}},
  \bibinfo{pages}{2865} (\bibinfo{year}{1998}).

\bibitem[{\citenamefont{Friedrich and Nagy}(1999)}]{FriedrichNagy1999}
\bibinfo{author}{\bibfnamefont{H.}~\bibnamefont{Friedrich}} \bibnamefont{and}
  \bibinfo{author}{\bibfnamefont{G.}~\bibnamefont{Nagy}},
  \bibinfo{journal}{Commun.\ Math.\ Phys.} \textbf{\bibinfo{volume}{201}},
  \bibinfo{pages}{619} (\bibinfo{year}{1999}).

\bibitem[{\citenamefont{Bardeen and Buchman}(2002)}]{Bardeen2002}
\bibinfo{author}{\bibfnamefont{J.~M.} \bibnamefont{Bardeen}} \bibnamefont{and}
  \bibinfo{author}{\bibfnamefont{L.~T.} \bibnamefont{Buchman}},
  \bibinfo{journal}{Phys.\ Rev.\ D} \textbf{\bibinfo{volume}{65}},
  \bibinfo{pages}{064037} (\bibinfo{year}{2002}).

\bibitem[{\citenamefont{Szil\'agyi et~al.}(2002)\citenamefont{Szil\'agyi,
  Schmidt, and Winicour}}]{Szilagyi2002}
\bibinfo{author}{\bibfnamefont{B.}~\bibnamefont{Szil\'agyi}},
  \bibinfo{author}{\bibfnamefont{B.}~\bibnamefont{Schmidt}}, \bibnamefont{and}
  \bibinfo{author}{\bibfnamefont{J.}~\bibnamefont{Winicour}},
  \bibinfo{journal}{Phys.\ Rev.\ D} \textbf{\bibinfo{volume}{65}},
  \bibinfo{pages}{064015} (\bibinfo{year}{2002}).

\bibitem[{\citenamefont{{Calabrese} et~al.}(2003)\citenamefont{{Calabrese},
  {Pullin}, {Reula}, {Sarbach}, and {Tiglio}}}]{Calabrese2003}
\bibinfo{author}{\bibfnamefont{G.}~\bibnamefont{{Calabrese}}},
  \bibinfo{author}{\bibfnamefont{J.}~\bibnamefont{{Pullin}}},
  \bibinfo{author}{\bibfnamefont{O.}~\bibnamefont{{Reula}}},
  \bibinfo{author}{\bibfnamefont{O.}~\bibnamefont{{Sarbach}}},
  \bibnamefont{and} \bibinfo{author}{\bibfnamefont{M.}~\bibnamefont{{Tiglio}}},
  \bibinfo{journal}{Commun.\ Math.\ Phys.} \textbf{\bibinfo{volume}{240}},
  \bibinfo{pages}{377} (\bibinfo{year}{2003}), \eprint{gr-qc/0209017}.

\bibitem[{\citenamefont{Szil\'agyi and Winicour}(2003)}]{Szilagyi2003}
\bibinfo{author}{\bibfnamefont{B.}~\bibnamefont{Szil\'agyi}} \bibnamefont{and}
  \bibinfo{author}{\bibfnamefont{J.}~\bibnamefont{Winicour}},
  \bibinfo{journal}{Phys.\ Rev.\ D} \textbf{\bibinfo{volume}{68}},
  \bibinfo{pages}{041501(R)} (\bibinfo{year}{2003}).

\bibitem[{\citenamefont{Kidder et~al.}(2005)\citenamefont{Kidder, Lindblom,
  Scheel, Buchman, and Pfeiffer}}]{Kidder2005}
\bibinfo{author}{\bibfnamefont{L.~E.} \bibnamefont{Kidder}},
  \bibinfo{author}{\bibfnamefont{L.}~\bibnamefont{Lindblom}},
  \bibinfo{author}{\bibfnamefont{M.~A.} \bibnamefont{Scheel}},
  \bibinfo{author}{\bibfnamefont{L.~T.} \bibnamefont{Buchman}},
  \bibnamefont{and} \bibinfo{author}{\bibfnamefont{H.~P.}
  \bibnamefont{Pfeiffer}}, \bibinfo{journal}{Phys.\ Rev.\ D}
  \textbf{\bibinfo{volume}{71}}, \bibinfo{pages}{064020}
  (\bibinfo{year}{2005}).

\bibitem[{\citenamefont{Buchman and Sarbach}(2006)}]{Buchman2006}
\bibinfo{author}{\bibfnamefont{L.~T.} \bibnamefont{Buchman}} \bibnamefont{and}
  \bibinfo{author}{\bibfnamefont{O.~C.~A.} \bibnamefont{Sarbach}},
  \bibinfo{journal}{Class.\ Quantum Grav.} \textbf{\bibinfo{volume}{23}},
  \bibinfo{pages}{6709} (\bibinfo{year}{2006}).

\bibitem[{\citenamefont{Buchman and Sarbach}(2007)}]{Buchman2007}
\bibinfo{author}{\bibfnamefont{L.~T.} \bibnamefont{Buchman}} \bibnamefont{and}
  \bibinfo{author}{\bibfnamefont{O.~C.~A.} \bibnamefont{Sarbach}},
  \bibinfo{journal}{Class.\ Quantum Grav.} \textbf{\bibinfo{volume}{24}},
  \bibinfo{pages}{S307} (\bibinfo{year}{2007}).

\bibitem[{\citenamefont{Gottlieb and Hesthaven}(2001)}]{Gottlieb2001}
\bibinfo{author}{\bibfnamefont{D.}~\bibnamefont{Gottlieb}} \bibnamefont{and}
  \bibinfo{author}{\bibfnamefont{J.~S.} \bibnamefont{Hesthaven}},
  \bibinfo{journal}{J. Comput. Appl. Math.} \textbf{\bibinfo{volume}{128}},
  \bibinfo{pages}{83} (\bibinfo{year}{2001}), ISSN \bibinfo{issn}{0377-0427}.

\bibitem[{\citenamefont{Hesthaven}(2000)}]{Hesthaven2000}
\bibinfo{author}{\bibfnamefont{J.~S.} \bibnamefont{Hesthaven}},
  \bibinfo{journal}{Appl. Num. Math.} \textbf{\bibinfo{volume}{33}},
  \bibinfo{pages}{23} (\bibinfo{year}{2000}).

\bibitem[{\citenamefont{Ossokine et~al.}()\citenamefont{Ossokine, Kidder, and
  Pfeiffer}}]{Ossokine-etal-in-prep}
\bibinfo{author}{\bibfnamefont{S.}~\bibnamefont{Ossokine}},
  \bibinfo{author}{\bibfnamefont{L.~E.} \bibnamefont{Kidder}},
  \bibnamefont{and} \bibinfo{author}{\bibfnamefont{H.~P.}
  \bibnamefont{Pfeiffer}}, \emph{\bibinfo{title}{Precession--tracking
  coordinates for binary black hole simulations}}, \bibinfo{note}{in
  preparation}.

\bibitem[{\citenamefont{Szilagyi et~al.}(2009)\citenamefont{Szilagyi, Lindblom,
  and Scheel}}]{Szilagyi:2009qz}
\bibinfo{author}{\bibfnamefont{B.}~\bibnamefont{Szilagyi}},
  \bibinfo{author}{\bibfnamefont{L.}~\bibnamefont{Lindblom}}, \bibnamefont{and}
  \bibinfo{author}{\bibfnamefont{M.~A.} \bibnamefont{Scheel}},
  \bibinfo{journal}{Phys.\ Rev.\ D} \textbf{\bibinfo{volume}{80}},
  \bibinfo{pages}{124010} (\bibinfo{year}{2009}), \eprint{0909.3557}.

\bibitem[{\citenamefont{Kidder}(1995)}]{kidder95}
\bibinfo{author}{\bibfnamefont{L.~E.} \bibnamefont{Kidder}},
  \bibinfo{journal}{Phys.\ Rev.\ D} \textbf{\bibinfo{volume}{52}},
  \bibinfo{pages}{821} (\bibinfo{year}{1995}).

\bibitem[{\citenamefont{Garcia et~al.}(2012)\citenamefont{Garcia, Lovelace,
  Kidder, Boyle, Teukolsky et~al.}}]{Garcia:2012dc}
\bibinfo{author}{\bibfnamefont{B.}~\bibnamefont{Garcia}},
  \bibinfo{author}{\bibfnamefont{G.}~\bibnamefont{Lovelace}},
  \bibinfo{author}{\bibfnamefont{L.~E.} \bibnamefont{Kidder}},
  \bibinfo{author}{\bibfnamefont{M.}~\bibnamefont{Boyle}},
  \bibinfo{author}{\bibfnamefont{S.~A.} \bibnamefont{Teukolsky}},
  \bibnamefont{et~al.} (\bibinfo{year}{2012}), \eprint{1206.2943}.

\bibitem[{\citenamefont{Lovelace et~al.}(2008)\citenamefont{Lovelace, Owen,
  Pfeiffer, and Chu}}]{Lovelace2008}
\bibinfo{author}{\bibfnamefont{G.}~\bibnamefont{Lovelace}},
  \bibinfo{author}{\bibfnamefont{R.}~\bibnamefont{Owen}},
  \bibinfo{author}{\bibfnamefont{H.~P.} \bibnamefont{Pfeiffer}},
  \bibnamefont{and} \bibinfo{author}{\bibfnamefont{T.}~\bibnamefont{Chu}},
  \bibinfo{journal}{Phys.\ Rev.\ D} \textbf{\bibinfo{volume}{78}},
  \bibinfo{pages}{084017} (\bibinfo{year}{2008}).

\bibitem[{\citenamefont{Lovelace}(2009)}]{Lovelace2009}
\bibinfo{author}{\bibfnamefont{G.}~\bibnamefont{Lovelace}},
  \bibinfo{journal}{Class.\ Quantum Grav.} \textbf{\bibinfo{volume}{26}},
  \bibinfo{pages}{114002} (\bibinfo{year}{2009}).

\end{thebibliography}

\end{document}